\documentclass[fleqn,usenatbib,useAMS]{mnras}

\usepackage{newtxtext,newtxmath}
\usepackage[capitalize,nameinlink]{cleveref}

\usepackage[T1]{fontenc}
\usepackage{comment}
\usepackage{booktabs}

\usepackage{graphicx}	
\usepackage{amsmath}	
\usepackage{orcidlink}

\newcommand{\fid}{\textsf{FID}}
\newcommand{\noiso}{\textsf{NULL}}
\newcommand{\inv}{\textsf{INV}}

\title[Ripples in the baryon to dark matter ratio in $\Lambda$CDM]{Ripples in the baryon to dark matter ratio in $\Lambda$CDM: \\ implications for galaxy formation}

\author[O. Jessop et al.]{
Owen Jessop\orcidlink{0009-0005-5161-3119}$^{1,\,2}$\thanks{E-mail: owen.jessop2@durham.ac.uk},
Adrian Jenkins\orcidlink{0000-0003-4389-2232}$^{1}$,
Andrew Pontzen\orcidlink{0000-0001-9546-3849}$^{1}$,
Joop Schaye\orcidlink{0000-0002-0668-5560}$^{3}$,
Matthieu Schaller\orcidlink{0000-0003-4389-2232}$^{3,\,4}$,
\newauthor
and John C.\ Helly\orcidlink{0000-0002-2395-4902}$^{1}$
\\
$^{1}$Institute for Computational Cosmology, Department of Physics, Durham University, South Road, Durham, DH1 3LE, United Kingdom \\
$^{2}$Centre for Extragalactic Astronomy, Department of Physics, Durham University, South Road, Durham, DH1 3LE, United Kingdom \\
$^{3}$Leiden Observatory, Leiden University, PO Box 9513, 2300 RA Leiden, the Netherlands \\
$^{4}$Lorentz Institute for Theoretical Physics, Leiden University, PO box 9506, 2300 RA Leiden, the Netherlands
}

\date{Accepted XXX. Received YYY; in original form ZZZ}

\pubyear{\the\year{}}

\begin{document}
\label{firstpage}
\pagerange{\pageref{firstpage}--\pageref{lastpage}}
\maketitle

\begin{abstract}
We use the \texttt{FLAMINGO} galaxy formation model to quantify the impact of baryon–CDM isocurvature perturbations on galaxy formation in $\Lambda$CDM. In linear theory, these perturbations represent local, compensated variations in the ratio between the baryon and CDM densities; they freeze in amplitude at late times, with an RMS amplitude of $1.5\%$ on the Lagrangian scale of a $10^{11}\,\rm M_\odot$ halo ($0.85\, \rm{Mpc}$). Although such perturbations arise naturally within $\Lambda$CDM, most cosmological simulations and semi-analytic models to date omit them. These perturbations are strongly anti-correlated with the matter overdensity field such that halos form with baryon fractions below the cosmic mean, with earlier-collapsing halos exhibiting stronger baryonic suppression. To isolate the galaxy response, we analyse three hydrodynamical simulations with identical initial matter overdensity fields that: i) include isocurvature modes, ii) omit them, or iii) invert their amplitude. At $z=8$, isocurvature perturbations reduce the mean baryon fraction and star formation rates of resolved halos by $ 5\%$ and $ 12\%$, respectively, relative to the null-isocurvature case. These effects show no systematic dependence on halo mass and diminish steadily with time, reaching $ 0.1\%$ and $ 1\%$ by $z=0$. We develop a model based on spherical collapse that accurately reproduces the mean baryon fraction suppression. As high-redshift observations become increasingly routine, incorporating isocurvature perturbations into simulations and semi-analytic models will be important for robust predictions of early galaxy and black hole formation in the \textit{JWST} era.
\end{abstract}




\section{Introduction}
\label{sec:intro}

Prior to recombination, baryons and photons formed a tightly coupled relativistic fluid that supported acoustic oscillations on sub-horizon scales \citep{Peebles1970}. These oscillations suppressed the growth of baryonic perturbations on scales below the sound horizon, in contrast to cold dark matter (CDM), which grew unimpeded via gravity after matter–radiation equality (e.g. \citealt{Eisenstein1998, Coles2002}). Shortly after recombination, the baryon and photon fluids decoupled, imprinting the acoustic oscillations in the matter distribution as the BAO signal observed in the late-time galaxy distribution (e.g. \citealt{Eisenstein2005, Cole2005}), and in the anisotropies of the CMB (e.g. \citealt{Spergel2003, Planck2020}). This decoupling left residual differences between the baryon and CDM density and velocity fields (e.g. \citealt{Bond1984, Bardeen1986}). The relative velocities, known as baryon–CDM streaming velocities, are coherent up to the BAO scale and remain supersonic with respect to the baryon sound speed after decoupling (e.g. \citealt{Tseliakhovich2010}). The accompanying relative density perturbations, known as baryon–CDM isocurvature perturbations, correspond to local, compensated variations in the ratio between the baryon and CDM densities. Here, a local baryon deficit is exactly offset by a CDM excess, and vice versa, such that the matter overdensity, and therefore the gravitational potential and spacetime curvature, remains unchanged while power is redistributed between the two components. These perturbations sourced by baryon–photon decoupling are fundamentally distinct from the inflationary compensated isocurvature perturbations discussed in other studies (e.g. \citealt{Heinrich2019, Barreira2020b}), which arise in multi-field inflationary models and are not a generic prediction of $\Lambda$CDM.

After decoupling, baryons began to “catch up” by falling into the gravitational potential wells established by CDM. Initially, the baryon overdensity field exhibited negligible small-scale structure relative to CDM due to its prior coupling to photons. As gravitational infall proceeded, the clustering of the two components gradually converged: the baryon power spectrum was suppressed by $\approx 50\%$ by $z=100$, $\approx 20\%$ by $z=30$, and $\approx 10\%$ by $z=10$. By the present day, these differences have largely been erased, with the baryon and CDM power spectra differing by $\lessapprox 1\%$ on scales above the Jeans length (e.g. \citealt{Angulo2013}). In contrast, the absolute amplitude of the baryon–CDM isocurvature overdensity field remains effectively frozen in amplitude at late times (e.g. \citealt{Naoz2007, Khoraminezhad2021, Hahn2021}), preserving a memory of the pre-decoupling baryon–photon oscillations. At late times, matter perturbations continue to grow whereas isocurvature modes remain approximately fixed in amplitude, rendering their influence most significant during the earliest phases of non-linear structure formation. This naturally raises the question of how strongly such perturbations can modulate early galaxy formation.

High-redshift galaxy formation has been revolutionised by the launch of the James Webb Space Telescope (\textit{JWST}; \citealt{JWST2023}), which has opened an observational window onto a previously inaccessible regime and discovered numerous bright galaxies at $z\gtrsim10$. Over the coming decade, ground-based facilities such as the Extremely Large Telescope (\textit{ELT}; \citealt{Padovani2023}), and the Giant Magellan Telescope (\textit{GMT}; \citealt{Kautz2024}) will complement \textit{JWST} with deep, high-resolution spectroscopy and spatially resolved observations of the high-redshift galaxy population. Together, these facilities will significantly deepen our understanding of galaxy formation in the first billion years after the Big Bang, thereby placing increasingly stringent demands on theoretical models to reproduce the emerging high-redshift landscape.

These observations are typically interpreted through comparisons with predictions from cosmological simulations of galaxy formation. However, historically many simulations treated baryons and CDM as tracers of a single effective mass-weighted baryon+CDM (matter) fluid in the initial conditions (ICs), the so-called one-fluid approximation. This was motivated by the difficulty of generating accurate two-fluid baryon--CDM ICs (e.g. \citealt{Angulo2013, Valkenburg2017, Hahn2021}), and by sizeable uncertainties in subgrid prescriptions that typically dwarfed any modest inaccuracies in the initial baryon–CDM distribution. Moreover, flagship simulation suites, such as \texttt{OWLS} \citep{Schaye2010}, \texttt{EAGLE} \citep{Schaye2015}, \texttt{BAHAMAS} \citep{McCarthy2017}, \texttt{IllustrisTNG} \citep{Pillepich2017}, \texttt{HORIZON-AGN} \citep{Kaviraj2017}, \texttt{SIMBA} \citep{Dave2019}, and \texttt{MillenniumTNG} \citep{Pakmor2023}, were designed to target low-redshift galaxy populations, where observational constraints are strongest and the effects of isocurvature perturbations are expected to be smallest. However, the landscape has been shifting in recent years: feedback models are becoming increasingly sophisticated and attention has been turning toward galaxy formation during the epoch of reionisation and beyond, as exemplified by simulation suites such as \texttt{SPHINX} \citep{Rosdahl2018}, \texttt{FLARES} \citep{Lovell2020}, \texttt{THESAN} \citep{Kannan2021}, \texttt{SERRA} \citep{Pallottini2022}, and \texttt{MEGATRON} \citep{Katz2025}. In this regime, the accuracy of the ICs becomes important, since neglecting the complete two-fluid baryon-CDM dynamics may introduce systematic biases in predictions of high-redshift observables.

Recent work by \citet{Rampf2020} and \citet{Hahn2021} has enabled the generation of higher-order two-fluid ICs that accurately reproduce the differential growth between baryons and CDM. These advances are implemented in the publicly available \texttt{MonofonIC} code\footnote{https://bitbucket.org/ohahn/monofonic} \citep{Michaux2020,Hahn2021}, which can generate up to third-order ICs for both Lagrangian and Eulerian simulations. The \texttt{FLAMINGO} \citep{Schaye2023,Kugel2023} and \texttt{COLIBRE} \citep{Schaye2025, Chaikin2025a} simulation suites provided the first implementation of these improved two-fluid ICs within hydrodynamical simulations, but the impact of the two-fluid treatment has not been assessed within these models. In this study, we present a dedicated suite of three simulations using the \texttt{FLAMINGO} galaxy formation model to quantify the impact of baryon–CDM isocurvature perturbations on galaxy formation.

\begin{figure*}
  \centering
  \includegraphics[width=\textwidth]{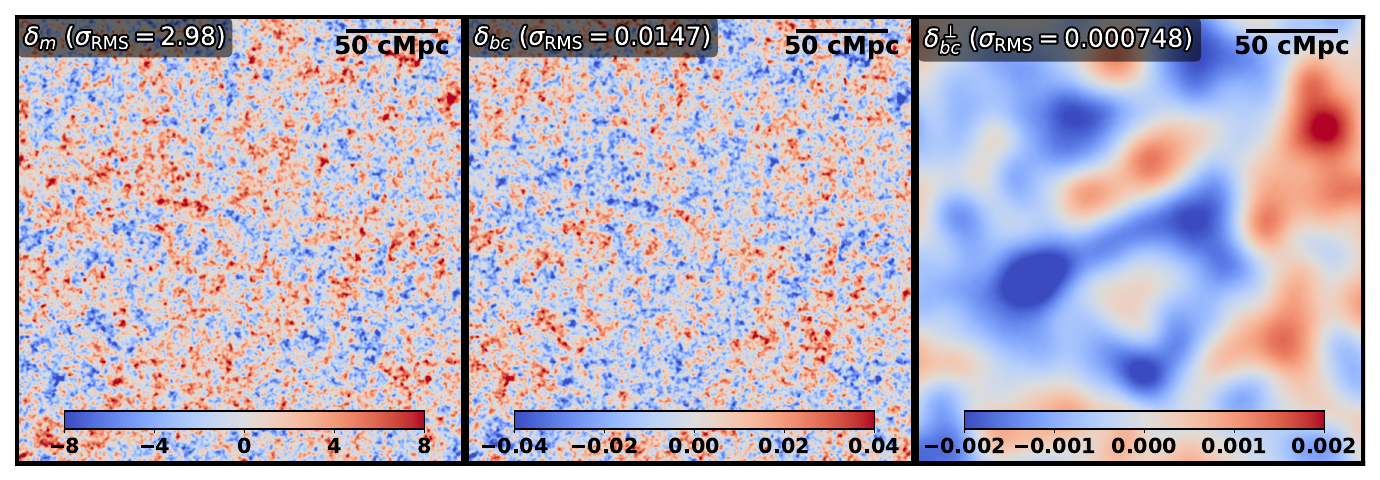}
  \caption{
    $z=0$ linear density fields computed with \texttt{CLASS} for the matter overdensity field, $\delta_{\rm m}$, the isocurvature overdensity field, $\delta_{\rm{bc}}$, and the orthogonal isocurvature overdensity field, $\delta_{\rm{bc}}^{\perp}$, see \cref{eq:delta_m1,eq:delta_bc2,eq:decomp3}. The fields have been smoothed with a spherical top-hat filter on the Lagrangian scale corresponding to a $10^{11}\,\rm{M_\odot}$ halo ($R =0.85\,\rm{Mpc}$). The cosmological parameters used correspond to the fiducial cosmology adopted throughout this paper (see \cref{subsec:flamingo_model}). We sample the fields on a periodic $500\,\rm{Mpc}$ comoving cubic volume with $900^3$ voxels; we display the central $250\,\rm{Mpc}$ region. Each panel shows an $x$–$y$ slice through the box centre with the root-mean-square (RMS) amplitude, $\sigma_{\rm{RMS}}$, given in the upper left corner of each panel.
}
  \label{fig:density_fields}
\end{figure*}

Most simulation-based work to date has focused on the baryon–CDM streaming velocity and its consequences for early galaxy formation. These studies show that a non-zero relative velocity delays the collapse of the first halos, suppresses the abundance of low-mass halos, reduces baryon fractions at high redshift, and postpones the onset of Population III star formation (e.g. \citealt{Naoz2012,Fialkov2014,Popa2016,Blazek2016,Schauer2019,Munoz2019,Park2020,Schauer2023}). In contrast, the role of isocurvature density perturbations has been explored less thoroughly in simulations. Existing work on these perturbations is dominated by analytic or perturbative treatments that predict, for example, shifts in the BAO peak, modifications to the galaxy power spectrum, and a suppression of halo gas content, among other effects (e.g. \citealt{Naoz2005,Barkana2005,Naoz2007,Barkana2011,Schmidt2016,Chen2019}). However, work exploring their effects in cosmological simulations, which is important for quantifying their impact on galaxy formation, remains relatively sparse and underdeveloped.

One of the earliest simulation-based comparisons of one-fluid and two-fluid ICs was carried out by \citet{Naoz2011}. Using very small ($2\,\textrm{Mpc}$) hydrodynamical simulations performed with \texttt{GADGET-2} \citep{Springel2005c}, they quantified how two-fluid ICs modify the filtering mass and the degree of gas suppression at $z \geq 11$ relative to the one-fluid case, finding a pronounced redshift dependence in both quantities. However, their ICs also incorporated spatial variations in the baryonic sound speed (e.g. \citealt{Naoz2005}). For the low-mass, high-redshift halos they targeted, this additional effect is quite significant, making it difficult to isolate the influence of isocurvature density perturbations, which is our primary interest.

Only in recent years have isocurvature perturbations been investigated in large-volume cosmological simulations. \citet{Barreira2020} introduced a long-wavelength isocurvature mode using the separate-Universe formalism,\footnote{The separate-Universe ansatz states that local structure formation within a long-wavelength perturbation in some fiducial cosmology is equivalent to global structure formation in an appropriately modified cosmology.} globally reducing the mean baryon mass fraction by $5\%$ across the simulation volume. Evolving this setup with the \textsc{IllustrisTNG} model \citep{Pillepich2017} and comparing to a control run, they measured a negative halo-bias parameter for the isocurvature component whose magnitude increases with halo mass and redshift, a $5\%$ offset in the stellar mass–halo mass relation at low redshift, and sub-percent shifts in the galaxy power spectrum. As we will discuss, however, the strong anti-correlation between the isocurvature and matter overdensity fields complicates the interpretation of these results. \citet{Khoraminezhad2021} moved closer to a first-principles treatment by performing two-fluid gravity-only simulations with separate baryon and CDM fluids in the ICs. Their halo-bias measurements were broadly consistent with \citet{Barreira2020}, and they found that halo gas fractions at $z=0$ differ by $\lessapprox 1\%$ between dual- and single-fluid ICs. \citet{Bird2020} also adopted two-fluid ICs and evolved them with the galaxy formation model implemented in the \textsc{MP-Gadget} code \citep{MPGadget}. Their analysis focused on the one-dimensional Lyman-$\alpha$ forest power spectrum, where they identified $\approx10\%$ deviations relative to single-fluid ICs, but they did not explicitly investigate the consequence of isocurvature perturbations for galaxy formation. 

To our knowledge, no previous study has combined two-fluid ICs with an advanced galaxy formation model in a large-volume hydrodynamical simulation to quantify how density isocurvature perturbations influence galaxy formation and propagate into late-time galaxy properties.

The structure of this paper is as follows. In \cref{sec:correlation}, we introduce baryon–CDM isocurvature perturbations, explain their origin, describe their key properties, and examine their implications for galaxy formation. In \cref{subsec:flamingo_model} we provide a concise overview of the \texttt{FLAMINGO} galaxy formation model. In \cref{subsec:ics}, we describe the generation of the \texttt{FLAMINGO} ICs and the modifications introduced for the simulations used in this work. \cref{subsec:halo_finding} then outlines the method for the identification of (sub)halos and the software used to compute their properties. The main results are presented in \cref{sec:results} where we consider the effects of isocurvature perturbations on the galaxy population. Finally, in \cref{sec:discussion}, we summarise our findings and discuss their broader implications for simulations of high-redshift galaxy formation.

\section{Baryon--CDM isocurvature perturbations in $\Lambda$CDM}
\label{sec:correlation}
After decoupling, baryons and CDM evolve jointly under gravity, effectively behaving as a single matter (baryon+CDM) fluid. However, the distinct pre-decoupling evolutionary histories leave  residual perturbations in the relative density of the two fluids, manifesting as baryon–CDM isocurvature modes.

The matter overdensity field, $\delta_{\rm m}(\mathbf{x},\,z)$, is defined as
\begin{equation}
\label{eq:delta_m1}
\delta_{\rm m}(\mathbf{x},\,z) \equiv \bar{f_{\rm b}}\, \delta_{\rm b}(\mathbf{x},\,z) + \bar{f_{\rm c}}\,\delta_{\rm c}(\mathbf{x},\,z),
\end{equation}
where $\mathbf{x}$ denotes the comoving position and $z$ the redshift. The baryon and CDM overdensities, $\delta_{\rm b}$ and $\delta_{\rm c}$, enter with weights given by their cosmic mean mass fractions, $\bar{f_{\rm b}}$ and $\bar{f_{\rm c}}$, which satisfy $\bar{f_{\rm b}} + \bar{f_{\rm c}} = 1$. These fractions are defined by
$\bar{f_{\rm b}} = \Omega_{\rm b}/\Omega_{\rm m}$ and $\bar{f_{\rm c}} = 1 - \bar{f_{\rm b}}$, where $\Omega_{\rm b}$ and $\Omega_{\rm m}$ are the present-day baryon and matter (baryons + CDM) density parameters, respectively. The matter overdensity field represents the mass-weighted combination of the two fluids and directly sources the gravitational potential to which both components respond, thereby determining the growth of cosmic structure. 

The isocurvature overdensity field, $\delta_{\rm{bc}}(\mathbf{x},\,z)$, is defined as (e.g. \citealt{Schmidt2016})
\begin{equation}
\label{eq:delta_bc2}
\delta_{\rm{bc}}(\mathbf{x},\,z) \equiv \delta_{\rm b}(\mathbf{x},\,z) - \delta_{\rm c}(\mathbf{x},\,z),
\end{equation}
which quantifies the relative density perturbation between baryons and CDM imprinted prior to decoupling. The field thus specifies whether a proto-halo region begins collapse with an enhanced or suppressed baryon fraction relative to the cosmic mean, and in this way $\delta_{\rm{bc}}$ modulates the baryonic reservoir available for galaxy formation once the region collapses non-linearly. In $\Lambda$CDM, linear-theory transfer functions imply a strong anti-correlation between $\delta_{\rm m}$ and $\delta_{\rm bc}$ on the mass scales relevant for early halo formation (e.g. \citealt{Naoz2007, Schmidt2016, Khoraminezhad2021}). The origin and implications of this anti-correlation will be examined further in the following subsections.

\subsection{Decomposition of the isocurvature overdensity field}
\label{subsec:decomp}

The strength of the anti-correlation between $\delta_{\rm{bc}}$ and $\delta_{\rm m}$ can be understood by explicitly decomposing $\delta_{\rm{bc}}$ into the component proportional to $\delta_{\rm m}$ and a residual orthogonal component, $\delta_{\rm{bc}}^{\rm \perp}$,
\begin{equation}
\label{eq:decomp3}
\delta_{\rm{bc}}(\mathbf{x},\,z) 
= \alpha(z)\,\delta_{\rm m}(\mathbf{x},\,z) \;+\; \delta_{\rm{bc}}^{\perp}(\mathbf{x},\,z),
\end{equation}
with the projection coefficient $\alpha(z)$ defined via the following inner product:
\begin{equation}
\label{eq:alpha_def}
\alpha(z) \equiv
\frac{\displaystyle \int \delta_{\rm{bc}}(\mathbf{x},\,z)\,\delta_{\rm m}(\mathbf{x},\,z)\, \rm d^3\mathbf{x}}
{\displaystyle \int \delta_{\rm m}^2(\mathbf{x},\,z)\,\rm d^3\mathbf{x}}.
\end{equation}

\cref{fig:density_fields} shows the linear density fields at $z=0$ for $\delta_{\rm m}$, $\delta_{\rm{bc}}$, and $\delta_{\rm{bc}}^{\rm \perp}$, calculated using the Einstein-Boltzmann code \texttt{CLASS}\footnote{http://class-code.net/} \citep{LesgourguesTram2011, Lesgourgues2011}. The fields have been smoothed with a spherical top-hat filter on the Lagrangian scale corresponding to a $10^{11}\,\rm M_\odot$ halo ($0.85 \, \textrm{Mpc}$). On this scale, the strong anti-correlation between $\delta_{\rm m}$ and $\delta_{\rm{bc}}$ is immediately evident: the two fields exhibit almost identical structures but with opposite, rescaled amplitudes. By contrast, the orthogonal component $\delta_{\rm{bc}}^{\perp}$ is a very low-amplitude, visibly smoother field with a coherence length comparable to the BAO scale. In the realisation shown, $\delta_{\rm{bc}}^{\perp}$ contributes only $4.9\%$ of $\sigma_{\textrm{RMS}}(\delta_{\rm{bc}})$ at $z=0$ on this scale. On proto-halo scales, the local amplitude of $\delta_{\rm bc}$ is therefore almost entirely determined by the local amplitude of $\delta_{\rm m}$.

The anti-correlation between the two fields originates in the pre-decoupling dynamics of the baryon and CDM fluids. Prior to decoupling, baryonic perturbations were strongly suppressed, leading to $\delta_{\rm m} \approx \bar{f_{\rm c}}\delta_{\rm c}$ and $\delta_{\rm{bc}} \approx -\delta_{\rm c}$, implying near-perfect anti-correlation. After decoupling, baryonic pressure becomes almost negligible, and the two fluids evolve at essentially the same rate above the Jeans scale\footnote{On very small scales, finite temperature and pressure effects continue to suppress baryonic perturbations even after decoupling.} under the influence of their shared gravitational potential: $\delta_{\rm m}$ follows the dominant growing mode, while $\delta_{\rm bc}$ comprises a linear combination of a constant mode and a decaying relative-velocity mode (e.g. \citealt{Hahn2021}). As the Universe expands, the relative-velocity mode is rapidly damped, eliminating any remaining differential growth between the fluids and leaving only the constant mode. The amplitude of $\delta_{\rm bc}$ therefore freezes out, asymptotically approaching a constant RMS amplitude of $ 1.5\%$ when smoothed on the Lagrangian scale corresponding to a $10^{11}\,\rm M_\odot$ halo\footnote{As with $\delta_{\rm m}$, in the absence of CDM free streaming, the RMS amplitude of $\delta_{\rm bc}$ diverges as smaller and smaller scales are included, so the field must be smoothed to obtain a physically meaningful value.}. For completeness, \cref{appendix:simple_argument} presents a derivation of this behaviour in an Einstein–de Sitter (EdS) Universe, showing that the late-time asymptotic behaviour of $\delta_{\rm bc}$ arises naturally from cosmic expansion and that its amplitude is fixed at decoupling, growing by a factor of approximately three between decoupling and the present day.

Deviations from exact pointwise anti-correlation in real space arise because baryon--photon coupling imprints pronounced BAO structure in $\delta_{\rm b}$ that is not expressed in $\delta_{\rm c}$, providing the physical origin of the orthogonal component. Despite its small amplitude, this component remains dynamically important: its coherence on BAO scales modulates the baryon fraction over tens of megaparsecs, correlates the assembly histories of halos in the same large-scale environment, and introduces scatter in $\delta_{\rm{bc}}$ at fixed $\delta_{\rm m}$. 

The orthogonal component directly sources the large-scale baryon–CDM streaming velocity, $\mathbf{v}_{\rm bc} \equiv \mathbf{v}_{\rm b} - \mathbf{v}_{\rm c}$, first identified as dynamically important by \citet{Tseliakhovich2010}. This velocity field acts as a coherent headwind that advects baryons relative to CDM potential wells. Although its amplitude decays over time, $\mathbf{v}_{\rm{bc}}$ retains large-scale coherence, delaying halo collapse, suppressing gas accretion onto shallow potential wells, and postponing Population~III star formation, effects that are most significant at high redshifts ($z \gtrsim 10$) and in low-mass halos ($M \lesssim 10^{7}\,\rm M_\odot$) (e.g. \citealt{Fialkov2014,Popa2016,Schauer2019,Schauer2023}).

\subsection{Predicting the effects of isocurvature perturbations on halo formation}
\label{subsec:linear_collapse}

In $\Lambda$CDM, $\delta_{\rm m}$ determines the gravitational potential and hence the initial phase of halo collapse, while the baryonic assembly that follows depends sensitively on the baryon fraction at collapse, which follows the cosmic mean but is locally modulated by $\delta_{\rm{bc}}$. Because $\delta_{\rm m}$ and $\delta_{\rm bc}$ are strongly anti-correlated on proto-halo scales (see \cref{subsec:decomp}), the earliest-collapsing, high-amplitude peaks in $\delta_{\rm m}$ correspond to the most baryon-deficient proto-halo regions, while later-collapsing, lower-amplitude peaks form from progressively less baryon-deficient regions. $\delta_{\rm bc}$ thus induces a coherent modulation of the initial baryon content that depends on collapse redshift. The orthogonal component of $\delta_{\rm bc}$ introduces additional large-scale correlated scatter at fixed $\delta_{\rm m}$, but this contribution is small, as shown in \cref{subsec:decomp}, and is likely subdominant for galaxy formation. The strength of this modulation can be quantified by evaluating $\delta_{\rm bc}$ at the time of collapse, since this fixes the halo’s initial baryon fraction. 

We begin by relating the change in baryon fraction to the amplitude of $\delta_{\rm bc}$. In linear theory, the change in the baryon fraction induced by isocurvature perturbations is
\begin{equation}
\label{eq:frac_change}
\frac{f_{\rm b}}{\bar{f_{\rm b}}} = 1+(1-\bar{f_{\rm b}})\,\delta_{\rm bc}.
\end{equation}
At late times, $\delta_{\rm bc}$ is effectively frozen in amplitude (e.g. \citealt{Hahn2021}), so the corresponding perturbation to the baryon fraction is likewise frozen and simply carried forward to the epoch of halo collapse. We now require a way to connect this linear-theory prediction to the baryon fraction of a halo collapsing at a given epoch. To do so, we adopt the spherical-collapse model \citep{Gunn1972, Press1974}, in which a proto-halo region collapses when its linearly extrapolated matter overdensity reaches $\delta_{\rm crit} \approx 1.686$. On proto-halo scales, we can relate $\delta_{\rm bc}$ and $\delta_{\rm m}$ using the decomposition established in \cref{subsec:decomp}, which shows that $\delta_{\rm bc}$ is well approximated by the linear projection of $\delta_{\rm m}$,
\begin{equation}
\label{eq:bc_m}
\delta_{\rm{bc}}(\mathbf{x},\,z)\;\approx\;\alpha(z)\,\delta_{\rm m}(\mathbf{x},\,z).
\end{equation}
While $\delta_{\rm m}$ grows with the linear growth factor $D(z)$, $\delta_{\rm bc}$ remains approximately constant at late times. We normalised $D(z)$ such that $D(z=0)=1$ and therefore
\begin{equation}
\label{eq:alpha}
\alpha(z) = \frac{\alpha_{\rm 0}}{D(z)}, \qquad \alpha_{\rm 0} \equiv \alpha(z=0).
\end{equation}
Evaluating \cref{eq:bc_m} and \cref{eq:alpha} at the collapse redshift $z_{\rm coll}$, and imposing the spherical-collapse condition $\delta_{\rm m}(z_{\rm coll}) = \delta_{\rm crit}$, yields
\begin{equation}
\label{eq:analytic_model}
\delta_{\rm bc}(z_{\rm coll}) = \frac{\delta_{\rm crit}\,\alpha_{\rm 0}}{D(z_{\rm coll})}.
\end{equation}
Finally, combining \cref{eq:analytic_model} with \cref{eq:frac_change}, we obtain a prediction for the change in baryon fraction of a proto-halo region collapsing at $z_{\rm coll}$,
\begin{equation}
\label{eq:final model}
\frac{f_{\rm b}}{\bar{f_{\rm b}}}\bigl(z_{\rm coll}\bigr) =  1+\frac{(1-\bar{f_{\rm b}})\,\delta_{\rm crit}\,\alpha_{\rm 0}}{D(z_{\rm coll})} \approx 1-\frac{0.0069}{D(z_{\rm coll})}.
\end{equation}
with $\alpha_{\rm 0} = -0.0049$ and $\bar{f_{\rm b}} = 0.160$ in the fiducial cosmology adopted in this paper (see \cref{subsec:flamingo_model}).

This expression links the collapse redshift directly to the isocurvature imprint on proto-halo scales: earlier collapse (smaller $D$) corresponds to more negative $\delta_{\rm bc}$ and therefore a larger initial baryon deficit. This result rests on linear theory and the idealised spherical-collapse model, and thus omits halo tri-axiality along with the full non-linear dynamics of gravitational collapse, hydrodynamics, and other baryonic processes. Even so, \cref{eq:final model} provides a clear baseline for comparison with the baryon-fraction changes measured in our simulations, enabling us to identify how non-linear collapse and baryonic physics amplify, diminish, or redistribute the linear imprint of $\delta_{\rm bc}$.

\citet{Barkana2011} proposed an alternative model for estimating the impact of isocurvature perturbations on halo formation. They used the quantity $r_{\rm LSS}$ (first introduced by \citealt{Naoz2007}), defined as the ratio of $\delta_{\rm b}$ to $\delta_{\rm m}$ evaluated in Fourier space. This quantity is computed using \texttt{CAMB} \citep{Lewis1999} as a function of redshift and evaluated at an intermediate-scale mode, \( k = 1\,/\rm (h\,Mpc) \), where it serves as a proxy for the isocurvature-induced suppression of halo gas content. When compared with the simulations of \citet{Naoz2011}, this model systematically under-predicted the gas suppression and required an empirical rescaling. Our prediction, based on the spherical-collapse model, can be extrapolated to the much higher redshifts considered by \citet{Naoz2011} and, for example, yields a gas suppression of $11\%$ at $z = 20$. Although this represents an improvement, it still underestimates the gas suppression measured in their simulations. However, a detailed direct comparison is not feasible because their ICs includes the additional effect of the spatially varying baryonic sound speed (e.g. \citealt{Naoz2005, Naoz2007}) which becomes important for the high-redshift ($z \geq 11$), low-mass ($M \sim 10^5\,\mathrm{M}_\odot$) halos they studied.

\section{The Simulations}
\label{sec:simulations}
\subsection{The \texttt{FLAMINGO} galaxy formation model}
\label{subsec:flamingo_model}

The \texttt{FLAMINGO} simulation suite\footnote{https://flamingo.strw.leidenuniv.nl/} (\textbf{F}ull-hydro \textbf{L}arge-scale structure simulations with \textbf{A}ll-sky \textbf{M}apping for the \textbf{I}nterpretation of \textbf{N}ext \textbf{G}eneration \textbf{O}bservations) \citep{Schaye2023, Kugel2023} comprises large-volume cosmological simulations calibrated to match the $z=0$ observed stellar mass function \citep{Driver2022} and the gas mass fractions of galaxy clusters \citep{Kugel2023}. The \texttt{FLAMINGO} suite was run using the open-source gravity, hydrodynamics, and galaxy formation code \texttt{SWIFT} (\textbf{S}PH \textbf{W}ith \textbf{I}nter-dependent \textbf{F}ine-grained \textbf{T}asking)\footnote{https://swiftsim.com} \citep{Schaller2024} with the \texttt{SPHENIX} Smoothed Particle Hydrodynamics (SPH) scheme \citep{Borrow2022}. Baryonic physics is modelled with subgrid prescriptions for radiative cooling and heating \citep{Ploeckinger2020}, star formation \citep{Schaye2008}, stellar mass loss \citep{Wiersma2009,Schaye2015}, supernova feedback energy \citep{DallaVecchia2008,Chaikin2022,Chaikin2023}, BH seeding and growth, and thermal feedback from active galactic nuclei (AGN) \citep{Springel2005b,Booth2009,Bahe2022}. All subgrid parameters were calibrated using Gaussian process emulators \citep{Kugel2023}.

Star formation in \texttt{FLAMINGO} follows the pressure-based subgrid model of \citet{Schaye2008}. Dense gas is placed on an effective equation of state intended to represent an unresolved multiphase interstellar medium and is stochastically converted into stars at a rate calibrated to reproduce the Kennicutt--Schmidt relation \citep{Kennicutt1998}. Gas particles that satisfy the required proper density, overdensity, and temperature criteria are eligible for one-to-one conversion into collisionless star particles, which inherit the full mass of their progenitor gas particles.

Black hole accretion is modelled using a modified Bondi--Hoyle accretion prescription, capped at the Eddington limit \citep[e.g.][]{Springel2005b}, with the accretion rate boosted at high gas densities to account for the unresolved Bondi radius and multi-phase interstellar medium \citep[e.g.][]{Booth2009}. Black holes are seeded by converting the densest gas particle in a halo into a collisionless black hole particle once the host halo exceeds \(M_{\rm halo} = 2.757 \times 10^{11}\,{\rm M_\odot}\), as identified by the \texttt{FoF} algorithm. As this criterion depends only on halo mass, it is insensitive to the presence of isocurvature perturbations. Newly seeded black holes are assigned an initial subgrid mass of \(10^{5}\,{\rm M_\odot}\).

We adopt the intermediate-resolution "m9" subgrid calibration of the \texttt{FLAMINGO} galaxy formation model, employing identical galaxy formation and feedback physics to those of the $2.8\,\rm{Gpc}$ flagship \texttt{FLAMINGO} run. The cosmological parameters are taken from the '3$\times$2pt + all external constraints' Dark Energy Survey Year 3 (DES-Y3) results \citep{Abbott2022}: $\Omega_{\rm m} = 0.306,\,\Omega_{\rm b} = 0.0486,\,\sigma_{\rm 8} = 0.807,\,\rm H_{\rm 0} = 68.1\,\rm{km\,s^{-1}\,Mpc^{-1}},\, \rm n_{\rm s} = 0.967$. 

The \texttt{FLAMINGO} simulations were initialised at $z_{\rm i}=31$ and run to a final redshift $z_{\rm f}=0$, with $z_{\rm i}$ chosen to be as late as possible to limit discreteness errors and reduce computational cost, while still preceding shell--crossing so that perturbation theory remains valid. 

For the simulations presented here, we adopt the same mass resolution as the \texttt{FLAMINGO} flagship run but evolve a smaller $(500\,\textrm{Mpc})^3$ volume, adjusting the particle counts accordingly. This configuration allows us to retain the m9 baryon particle mass resolution while keeping the computational cost tractable, and enables us to accurately resolve halos with $M \geq 10^{11}\,\rm{M_\odot}$. Our analysis uses 78 simulation snapshots spanning \(z=8\) to \(z=0\).

\subsection{The initial conditions}
\label{subsec:ics}

The generation of ICs for our simulations follows the same procedure as for the original \texttt{FLAMINGO} suite, whose ICs were generated with a modified version of \texttt{MonofonIC}\footnote{https://github.com/wullm/monofonic} \citep{Elbers2022} that self-consistently incorporates massive neutrinos alongside the two-fluid baryon–CDM formalism of \citet{Hahn2021}. Details of the neutrino implementation within \texttt{FLAMINGO} are provided in \citet{Schaye2023}. Following \citet{Rampf2020}, \texttt{MonofonIC} extends single-fluid Lagrangian perturbation theory (LPT) to a two-fluid system under the assumption that baryon pressure and temperature are negligible (i.e. a Jeans length much smaller than the scales of interest; e.g. \citealt{Hahn2021}). On sufficiently small scales this approximation formally breaks down, as $\Lambda$CDM predicts a scale-dependent baryonic sound speed sourced by spatial temperature fluctuations \citep{Naoz2005, Naoz2007}. However, these effects are expected to be negligible for the scales and redshifts analysed here. Simple estimates based on the Jeans mass indicate that baryonic pressure becomes dynamically relevant only for haloes with masses of order $10^{5}$--$10^{6}\, \rm M_\odot$, and detailed calculations and simulations are consistent with this range \citep[e.g.][]{Gnedin1998, Gnedin2000, Barkana2001,Naoz2009}. At the higher halo masses considered in this work, any associated temperature fluctuations can therefore be safely neglected.

In the framework of \citet{Hahn2021}, the baryon and CDM overdensity fields are described by three linear modes: the dominant adiabatic growing mode sourced by the total gravitational potential, and two baryon–CDM isocurvature modes, consisting of a constant mode and a decaying mode associated with the baryon–CDM streaming velocity. Since standard LPT only self-consistently treats the growing mode, \texttt{MonofonIC} first generates particle displacements from the dominant growing-mode solution and then encodes the constant mode via local perturbations to the particle masses. The magnitude of these mass perturbations is set by the local amplitude of $\delta_{\rm bc}$ evaluated at late times, typically $z=0$. The decaying mode is incorporated as a first-order correction to particle masses and velocities, thereby introducing the appropriate redshift dependence into the effective mass perturbations that encode $\delta_{\rm bc}$. This treatment is well justified at late times, for which the decaying mode has already been strongly suppressed and can be regarded as a small correction; in this regime, the approximation reproduces the linear-theory evolution of the baryon and CDM density fields to high accuracy.

In this study, we perform a suite of three simulation variants designed to isolate the impact of density isocurvature perturbations on galaxy formation. For simplicity we omit the decaying mode---convergence tests in smaller-volume simulations (not shown) confirm that the relative-velocity contribution remains subdominant, since our simulations are not sensitive to the redshift or mass resolution ($M \lesssim10^{7}\,\rm M_{\odot}$ at $z\gtrsim10$) where streaming velocities significantly affect halo formation (e.g. \citealt{Fialkov2012, Popa2016, Schauer2019, Schauer2023}). All three simulations share an identical Gaussian white-noise field drawn from a subregion of \texttt{Panphasia} \citep{Jenkins2013}, ensuring that their initial matter overdensity fields are identical. The chosen white noise field matches that of the L2p8\_m9 \texttt{FLAMINGO} variant and the three simulations differ only in their initial particle masses, which determine the specific isocurvature modes:
\begin{itemize}
  \item \textbf{\fid{} (FIDUCIAL: Negatively correlated isocurvature)}:  
    Baseline run using the original \texttt{MonofonIC} particle masses, retaining isocurvature modes (apart from the small omitted decaying mode). This provides the closest realisation of $\Lambda$CDM.  
  \item \textbf{\noiso{} (NULL: No isocurvature)}:  
    Erases isocurvature by resetting baryon and CDM particle masses to their respective global mean values. Identical to the single-fluid approximation used in many previous studies, in which baryons and CDM are initialised using the matter overdensity field.
  \item \textbf{\inv{} (INVERSE: Positively correlated isocurvature)}:  
  Inverts the isocurvature amplitude by reflecting particle masses about their mean, creating isocurvature modes positively correlated with the matter overdensity. This is not a physical solution in $\Lambda$CDM, but constitutes a useful test case.
\end{itemize}

For the adopted cosmology and particle load, the mean particle masses are
\begin{equation}
\bar{m}_{\rm c} = (5.65\times10^{9})\,\rm M_\odot,\qquad
\bar{m}_{b} = (1.07\times10^{9})\,\rm M_\odot,
\end{equation}
for CDM and baryons, respectively. In the isocurvature simulations, the particle masses exhibit RMS fractional deviations of $1.82\%$ for baryons and $0.34\%$ for CDM, corresponding to the same absolute RMS mass deviation of $1.95\times10^{7}\,\rm M_\odot$ for both components. Because the particle-mass perturbations are compensated by construction, the matter overdensity field is unchanged by these modifications. All three simulations therefore begin from identical initial matter overdensity fields, ensuring that any subsequent differences in galaxy formation arise solely from the imposed baryon–CDM isocurvature modes.

\subsection{Structure \& substructure identification}
\label{subsec:halo_finding}

The identification of (sub)halos within our simulations is performed via a two‐step procedure. First, field halos are identified using a standard \texttt{FoF} algorithm (\textbf{F}riends‐\textbf{o}f‐\textbf{F}riends; \citealt{Press1982}), producing \texttt{FoF} catalogues for each snapshot. The \texttt{FoF} algorithm links all pairs of particles separated by less than $b$ times the mean inter‐particle separation, with $b=0.2$ \citep{Davis1985} adopted in this work, corresponding to an overdensity of approximately $100\,\rho_{\textrm{mean}}$ \citep{More2011}, where $\rho_{\textrm{mean}}$ denotes the mean density of the Universe. Only CDM particles are considered as linkable. Baryonic particles (gas, stars, BHs) cannot directly form links to one another; instead, they only form links with CDM particles.

Second, substructures within these \texttt{FoF} groups are identified with the \texttt{HBT-HERONS}\footnote{https://github.com/SWIFTSIM/HBT-HERONS} algorithm (\textbf{H}ierarchical \textbf{B}ound \textbf{T}racing – \textbf{H}ydro-\textbf{E}nabled \textbf{R}etrieval of \textbf{O}bjects in \textbf{N}umerical \textbf{S}imulations; \citealt{Moreno2025}). \texttt{HBT-HERONS} is an updated version of the history‐based subhalo finder \texttt{HBT+} \citep{Han2012, Han2017}, featuring improved identification and tracking of subhalos in both dark matter–only and hydrodynamical simulations.

The properties of the (sub)halos identified by \texttt{HBT-HERONS} are computed using \texttt{SOAP}\footnote{https://github.com/SWIFTSIM/SOAP} (\textbf{S}pherical \textbf{O}verdensity and \textbf{A}perture \textbf{P}rocessor; \citealt{McGibbon2025}). \texttt{SOAP} is a Python package that reads the (sub)halo centres and particle membership lists provided by the halo finder as input and calculates a variety of subhalo properties within different spherical and projected apertures from the particle data. 

In this work, we adopt the spherical overdensity definition of halos and compute halo properties including all particles within $R_{\rm 200}$, the radius enclosing a mean density of $200\rho_{\rm crit}$, where $\rho_{\rm crit}$ is the critical density of the Universe. To ensure that the halos in our analysis are sufficiently well resolved, we apply a minimum mass threshold of $10^{11}\,\rm{M_\odot}$.

\section{Results}
\label{sec:results}

In this section we quantify the impact of baryon–CDM isocurvature perturbations on galaxy formation by comparing global quantities across our three simulations, using the halo definition described in \cref{subsec:halo_finding}. We focus on global measures because no statistically significant halo-mass dependence was found at fixed redshift. Uncertainties on global quantities are estimated via jackknife resampling, in which each simulation volume is divided into eight subregions and each global quantity is recomputed while successively omitting one subregion, with the resulting variance capturing finite-volume and sample-variance effects. Although we present results for the \inv{} simulation, our primary focus is the comparison between \fid{} and \noiso{}, which directly tests the commonly adopted approximation of neglecting baryon–CDM density differences in the initial conditions; the \inv{} run instead provides a deliberately non-physical inversion of the isocurvature amplitude and serves as an internal consistency check. We restrict the analysis to redshifts $z \leq 8$, beyond which halo abundances become too sparse for robust interpretation.

\begin{figure}
  \centering
  \includegraphics[width=\columnwidth]{figures/baryon_mass.pdf}
  \caption{\textit{Top panel}: The time evolution of the comoving baryon mass density in resolved halos, $\rho_{\rm b}^{\rm halos}(z)$, for our three different simulations. \textit{Bottom panel}: The time evolution of the baryonic mass density in the \fid{} and \inv{} simulations relative to the \noiso{} simulation, expressed as a ratio. The dotted green lines in the bottom panel represent the predictions of the theoretical model derived in \cref{subsec:linear_collapse}, \cref{eq:final model}.}
  \label{fig:baryon_mass}
\end{figure}

\subsection{The time evolution of the baryon and gas content of halos}
\label{subsec:baryon_gas}

In \cref{subsec:linear_collapse}, we showed that linear theory combined with spherical collapse predicts an isocurvature-induced baryon deficit in halos relative to the null isocurvature case, arising solely from differences in their Lagrangian patches. We now test whether these initially small variations persist through non-linear structure formation by examining the time evolution of the comoving baryon mass density in resolved halos, $\rho_{\rm b}^{\rm halos}(z)$, shown in the top panel of \cref{fig:baryon_mass}.\ This quantity rises steeply from high redshift towards the present day; the first halos collapse from rare, high-amplitude peaks in the matter overdensity field, whereas lower-amplitude peaks that are far more numerous collapse later and dominate the mass budget at low redshift (e.g. \citealt{Fakhouri2010,Behroozi2013,Behroozi2019}). Accordingly, the baryon mass density in resolved halos grows by nearly three orders of magnitude between $z=8$ ($0.65\,\rm{Gyr}$ after the Big Bang) and $z=0$ ($ 13.76\,\rm{Gyr}$ after the Big Bang), before levelling off as the rate of structure formation declines at late times. By $z=0$, the incremental growth of $\rho_{\rm b}^{\rm halos}(z)$ is small, indicating that most of the baryons destined to reside in collapsed systems have already been incorporated into virialised halos, even though a substantial fraction of cosmic baryons remain in the diffuse IGM (e.g. \citealt{Peroux2020, Tuominen2021, Yang2022}).

The three simulations show nearly indistinguishable absolute values for $\rho_{\rm b}^{\rm halos}(z)$, particularly at low redshift, so to isolate the isocurvature response the bottom panel of \cref{fig:baryon_mass} displays $\rho_{\rm b}^{\rm halos}(z)$ in the \fid{} and \inv{} simulations expressed as ratios relative to the \noiso{} simulation. The dotted green lines represent the predictions of the theoretical spherical collapse model derived in \cref{subsec:linear_collapse}, \cref{eq:final model}. At $z=8$, the isocurvature variants deviate from \noiso{} by $ 5\%$, but this offset declines steadily to the sub-percent level by $z=0$. The theoretical prediction, \cref{eq:final model}, provides an accurate estimate of the isocurvature-induced suppression measured in the simulations. It successfully reproduces the overall trend, including the sign of the modulation and the slope of the relation.

\begin{figure*}
  \centering
  \includegraphics[width=\textwidth]{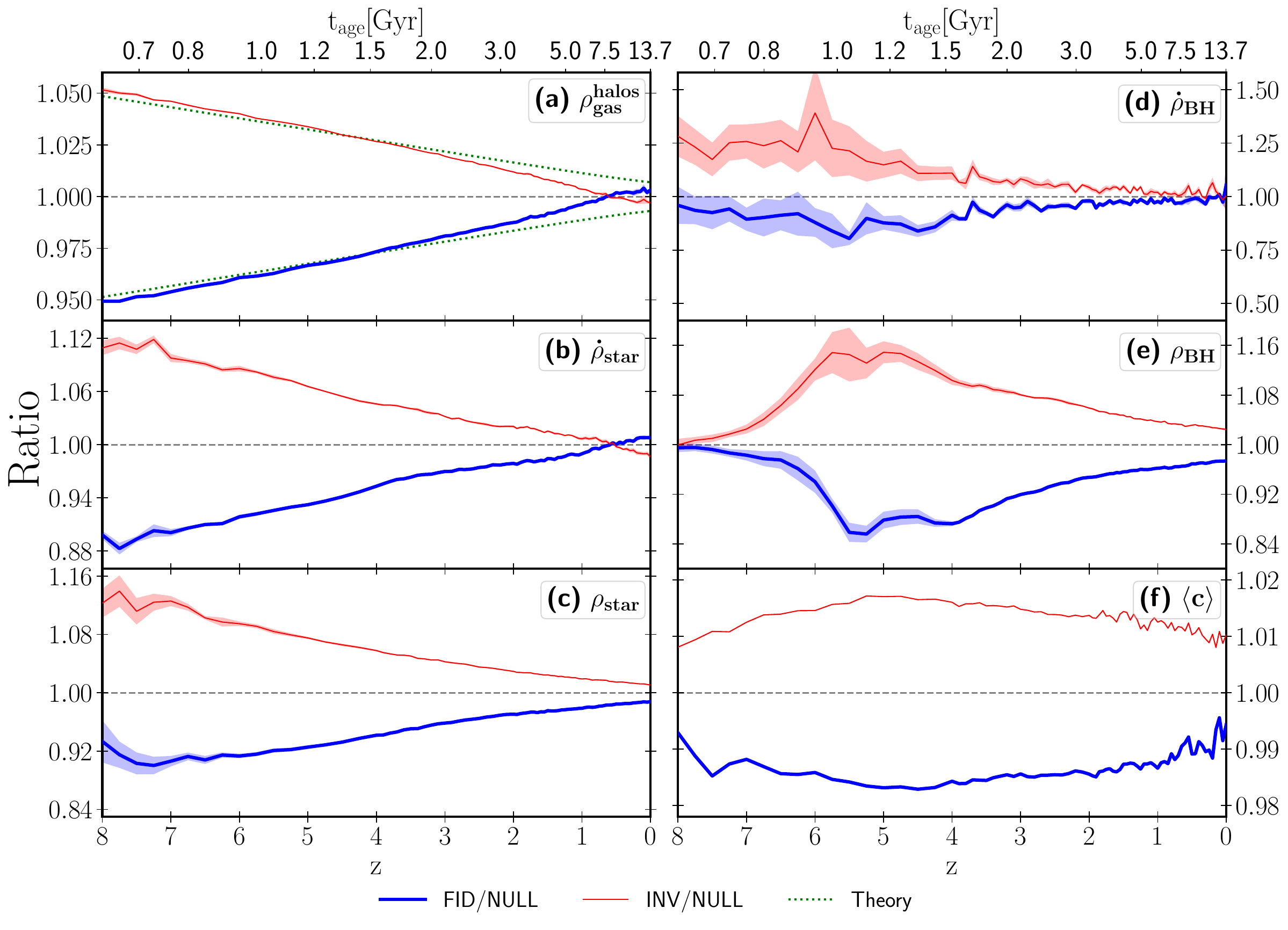}
  \caption{The time evolution of the ratio of various global quantities for the \fid{}/\noiso{} simulations (thick blue lines) and the \inv{}/\noiso{} simulations (thin red lines). The dotted green lines displayed in panel (a) represent the prediction of the theoretical model derived in \cref{subsec:linear_collapse}, \cref{eq:final model}. Jackknife uncertainties have been computed for each quantity, propagated into the corresponding ratios, and are shown as shaded regions. Panel (a): the comoving gas mass density in resolved halos, $\rho_{\rm gas}^{\rm halos}(z)$; Panel (b): the comoving star formation rate density, $\dot{\rho}_{\rm star}(z)$; Panel (c): the comoving stellar mass density, $\rho_{\rm star}(z)$; Panel (d): the comoving BH accretion rate density, $\dot{\rho}_{\rm{BH}}(z)$; Panel (e): the comoving BH mass density, $\rho_{\rm{BH}}(z)$; Panel (f): The mean halo concentration, $\langle c\rangle(z)$.} 
  \label{fig:volume_averaged}
\end{figure*}

Having established the time evolution of the baryonic reservoir, we now consider the gas component in isolation. For clarity, we focus on the ratio of the isocurvature simulations relative to \noiso{}, since our primary interest lies in the differential response between simulations. Panel (a) of \cref{fig:volume_averaged} shows the time evolution of the comoving gas mass density in resolved halos, $\rho_{\rm gas}^{\rm halos}(z)$. At early times, when the Universe is still dominated by cold, unprocessed gas, this quantity closely traces the comoving baryon mass density in resolved halos, $\rho_{\rm b}^{\rm halos}(z)$, since the majority of baryons remain in the gaseous phase. As time advances, radiative cooling and gravitational collapse progressively funnel gas into dense regions where it is converted into stars, accreted onto BHs, or expelled due to feedback. This transition causes the gas mass to deviate from the baryonic mass, and the theoretical model to break down at late times. 

The near-perfect antisymmetry of the \fid{} and \inv{} curves about the \noiso{} baseline further confirms that this behaviour is a direct consequence of the sign of the $\delta_{\rm{bc}}$–$\delta_{\rm m}$ (anti-)correlation. Positively and negatively correlated isocurvature overdensity fields impose opposing initial gas deficits and surpluses, which in turn generate lower or higher gas fractions in halos that would otherwise form from identical matter overdensities.

A particularly notable late-time feature is the crossover in the gas ratios: for $z \lesssim 0.7$, the \fid{} simulation exhibits a slightly higher collapsed gas mass than \noiso{}, while \inv{} falls modestly below it. The gradual convergence of the ratios toward unity can be partly explained by the fact that halos forming later are intrinsically less gas-deficient at formation; however, this effect alone cannot account for the reversal of the trend. The crossover instead arises from the coupled, time-dependent evolution of the gas reservoir, star formation, and feedback. In \fid{}, the initial gas deficit suppresses early star formation and BH fuelling, thereby reducing early feedback and gas consumption. The opposite effect occurs in \inv{}, where an initial gas surplus enhances early star formation and accretion activity, amplifying feedback and accelerating gas depletion.

\subsection{The time evolution of the stellar mass}
\label{subsec:stars}
The time evolution of the comoving star formation rate density, $\dot{\rho}_{\rm star}(z)$, shown in panel (b) of \cref{fig:volume_averaged}, quantifies the instantaneous rate at which gas is converted into stars at each epoch per unit volume. Its evolution reflects the complex interplay between the supply of cold, star-forming gas and the competition among three key processes: gas accretion from the intergalactic medium, which feeds the gas reservoir; radiative cooling and gravitational collapse within halo potentials, which enable star formation; and feedback from massive stars, supernovae, and active galactic nuclei (AGN), which heat or expel gas and thereby regulate subsequent star formation (e.g. \citealt{Madau2014}).

At $z = 8$, the \fid{} simulation exhibits a $12\%$ suppression of $\dot{\rho}_{\rm star}$ relative to \noiso{}, whereas the \inv{} simulation shows an approximately symmetric $12\%$ enhancement. These early-time deviations reflect the initial modulation of halo gas content by $\delta_{\rm{bc}}$, which either delays or accelerates the onset of cooling and star formation by altering the amount of gas available within each halo. The response of $\dot{\rho}_{\rm star}(z)$ is not directly proportional to that of $\rho_{\rm gas}^{\rm halos}(z)$. Instead, the star formation rate density exhibits offsets that are roughly twice as large, producing the late-time crossover seen in both quantities discussed in \cref{subsec:baryon_gas}. This amplified behaviour arises because star formation responds non-linearly to the local gas density (e.g. \citealt{Kennicutt1998}), and radiative cooling rates scale with the square of the gas density (e.g. \citealt{Smith2017, Lee2024}).

The integrated consequence of these differences is captured by the time evolution of the comoving stellar mass density, $\rho_{\rm star}(z)$, shown in panel (c) of \cref{fig:volume_averaged}. Here, the time evolution departs slightly from the near-perfect antisymmetry seen in the gas and star formation rates: the \inv{} simulation displays a $12\%$ enhancement in $\rho_{\rm star}(z)$, while the \fid{} run shows an $8\%$ suppression at $z = 8$. Although the large jackknife uncertainties reflect the small stellar mass formed by this epoch, the asymmetry appears physical.  It arises because the boosted gas supply in \inv{} not only increases the instantaneous star formation rate but also accelerates its onset, giving an early advantage. Since $\rho_{\rm star}(z)$ is closely related to the time-integrated star formation rate, even symmetric differences in $\dot{\rho}_{\rm star}$ do not lead to symmetric totals. This early imbalance diminishes as stellar mass builds up, and by $z = 5$ the isocurvature runs exhibit near-perfect symmetry about \noiso{}, with only a $1\%$ residual difference by $z = 0$. Thus, while isocurvature perturbations significantly modulate early star formation through their impact on gas content, their long-term influence on the integrated stellar mass budget is small within the \texttt{FLAMINGO} galaxy formation model.

\subsection{The time evolution of the black hole mass}
\label{subsec:black_holes}

The growth of black holes (BHs) provides an especially sensitive probe of isocurvature perturbations, owing to the highly non-linear coupling between gas supply, accretion physics, and feedback. Panel (d) of \cref{fig:volume_averaged} shows the time evolution of the comoving black hole accretion rate density, $\dot{\rho}_{\rm BH}(z)$, a global, population-averaged quantity that measures the instantaneous rate at which gas is accreted onto the entire BH population per unit comoving volume at each epoch. The accretion histories are distinctly asymmetric between the \inv{}/\noiso{} and \fid{}/\noiso{} models. Jackknife uncertainties are substantially larger, particularly at high redshift, reflecting the fact that the global accretion rate is dominated by a few very highly accreting BHs, consistent with other works (e.g. \citealt{Weinberger2025}). At $z=8$, \inv{} is enhanced by $25\%$ relative to \noiso{}, whereas \fid{} is mildly suppressed by $5\%$. This asymmetry arises from the coupling between the isocurvature-induced initial gas supply and the complex, non-linear growth and feedback cycles governing BH evolution, whereby modest initial differences in gas supply are non-linearly amplified.

These differing accretion histories are manifested in the time evolution of the comoving BH mass density, $\rho_{\rm BH}(z)$, shown in panel (e) of \cref{fig:volume_averaged}. At $z=8$, the simulations agree to within $\mathcal{O}(1\%)$, as expected from the \texttt{FLAMINGO} seeding prescription; because compensated isocurvature modes leave halo collapse times effectively unchanged, the number of seeded BHs at fixed redshift is nearly identical across all runs. At these early epochs, insufficient time has elapsed for differences in accretion histories to generate appreciable differences in the BH populations. The jackknife uncertainties on $\rho_{\rm BH}(z)$ are substantially smaller than those on $\dot{\rho}_{\rm BH}(z)$, consistent with the dominance of a few objects in setting the instantaneous global accretion rate.

Subsequently, $\rho_{\rm{BH}}$ diverges among the three simulations. In \inv{}, the ratio to \noiso{} rises rapidly and reaches a maximum offset of $15\%$ by $z=6$, driven by the larger accretion rate. The \fid{}/\noiso{} ratio shows a suppression of similar magnitude but evolves more gradually, peaking at $15\%$ at a later epoch, $z=5.5$. The near equality of these peak amplitudes is initially counter-intuitive given the strong asymmetry in $\dot{\rho}_{\rm{BH}}$. This arises because the BH population is growing extremely rapidly at early times: the comoving BH mass density increases by an order of magnitude over a short redshift interval, so even highly asymmetric instantaneous accretion rates integrate to a more symmetric cumulative effect. Thereafter the ratios evolve non-monotonically, briefly decreasing, then rising again before declining toward late times. By $z=0$ only a residual $3\%$ offset in $\rho_{\rm{BH}}$ remains between the isocurvature and \noiso{} runs. Among all diagnostics, the BH population thus retains the clearest late-time memory of isocurvature perturbations.

The greater symmetry in $\rho_{\rm{BH}}(z)$ compared with $\dot{\rho}_{\rm{BH}}(z)$ indicates that, despite the strongly asymmetric early accretion histories, the comoving BH mass density is regulated primarily by the sustained availability of gas. Although the global accretion rate at early times is dominated by a small number of BHs, the cumulative BH mass budget is distributed more evenly. Moreover, because the absolute accretion rates at high redshift are still small and increase steeply from snapshot to snapshot, the ratio of $\rho_{\rm{BH}}$ at any given epoch is set almost entirely by the accretion history in the short interval immediately preceding it.

\subsection{The time evolution of the mean halo concentration}
\label{subsection:halo concentration}

The halo concentration, c, is a dimensionless measure of the central mass distribution of a halo and provides a useful diagnostic of whether isocurvature perturbations induce structural changes beyond their impact on the baryonic mass in halos. Panel (f) of \cref{fig:volume_averaged} presents the ratio plots for the time evolution of the mean halo concentration, $\langle c \rangle(z)$, defined as the average value of $c_{\rm 200}$ across all halos in the simulation volume. The halo concentration is computed following the method outlined in \citet{Wang2023b}, but employing a fifth-order polynomial fit to the $R_{\rm 1}$–c relation over the range $1 < c < 1000$, where $R_{\rm 1}$ denotes the first moment of the density distribution, under the assumption that halos follow a Navarro–Frenk–White (NFW) density profile \citep{Navarro1997}. In the \fid{} simulation, isocurvature perturbations suppress the concentration by $1\%$ at $z=8$, rising to $1.5\%$ suppression at $z=5$, then returning to $1\%$ by $z=0$, with the \inv{} simulations exhibiting a nearly mirrored trend. Although one might expect stronger suppression of the gas content at high redshift to reduce cooling and adiabatic contraction, thereby producing a redshift-dependent decline in concentration that tracks the gas deficit, the observed near time-independence of the suppression suggests that the underlying behaviour is more complex. Given the small amplitude and weak redshift evolution, we do not pursue this further here and conclude that halo concentrations are largely insensitive to the presence of isocurvature perturbations.

\section{Discussion \& Conclusions}
\label{sec:discussion}
    
In this work we have presented a suite of three cosmological simulations performed with the \texttt{FLAMINGO} galaxy‐formation model to quantify the impact of baryon–CDM isocurvature perturbations on key global galaxy properties. Introducing isocurvature perturbations into the ICs results in an average suppression of $5\%$ in the comoving gas/baryon mass density of resolved halos at $z=8$. The simulations show excellent agreement with the theoretical model developed in \cref{subsec:linear_collapse}, which predicts the mean baryon suppression in halos induced by isocurvature perturbations. The reduced gas supply yields a $12\%$ suppression in the comoving star formation rate density and a comparable reduction in the comoving stellar mass density. The response of the comoving BH mass density lags the gas supply, with a maximum suppression of $15\%$ at $z=5.5$. The mean halo concentration exhibits a small response to isocurvature perturbations, with a suppression of just $1-2\%$ that is weakly redshift dependent. These quantitative predictions should be interpreted in light of the scope of the \texttt{FLAMINGO} galaxy formation model, which was designed for studying structure formation in large cosmological volumes, at the expense of mass resolution and the detailed modelling of individual galaxies. The results reported here, other than the comoving baryon mass density in resolved halos, are therefore likely to be subgrid-model dependent. 

Although the corrections introduced due to isocurvature perturbations are modest compared with present observational and subgrid-model uncertainties, their relative importance increases toward high redshift. The theoretical model presented in \cref{subsec:linear_collapse} predicts a $9\%$ reduction in the gas content of halos collapsing at $z=15$, which suggests a suppression of the comoving star formation rate density by over $20\%$ at that epoch. 

The high-redshift Universe at $z \geq 10$, until recently inaccessible, is now being probed by \textit{JWST} \citep{JWST2023} surveys such as \textit{JADES} \citep{Bunker2019} and \textit{CEERS} \citep{Finkelstein2023}, which have reported unexpectedly luminous galaxies within the first few hundred million years after the Big Bang. These findings have prompted debate over whether they challenge $\Lambda$CDM or instead reflect limitations in current galaxy formation models (e.g.\ \citealt{Menci2022,Wang2023a,McCaffrey2023,Gupta2023,Lu2024,Stark2025}). Our analysis shows that isocurvature perturbations induce an $\mathcal{O}(10\%)$ suppression in stellar mass and star formation rate densities at $z=8$, with the effect increasing toward higher redshift. Although modest relative to present theoretical uncertainties, this systematic shift acts counter to the observed \textit{JWST} excess, predicting fewer bright galaxies at early times and thereby marginally worsening the apparent tension. As observations continue to improve, accounting for these physically motivated corrections will be essential for robust comparisons between simulations, semi-analytic models, and data. 

Another tension highlighted by \textit{JWST} concerns the existence of numerous supermassive black holes (SMBHs) at high redshift. Deep observational campaigns have discovered quasars at $z>8$ hosting SMBHs with masses approaching $10^9\,\rm M_\odot$ (e.g.\ \citealt{Juodzbalis2023, Goulding2023, Maiolino2024, Pacucci2024, Juodbalis2024}), as well as a growing population of compact, extremely red sources, known as `little red dots', whose morphologies, colours, and spectral signatures are consistent with actively accreting SMBHs (e.g.\ \citealt{Labbe2023, Matthee2024, Kocevski2025}). Taken together, these observations imply that black holes were assembling mass at extreme rates within the first few hundred million years after the Big Bang. Although the inferred masses and accretion rates remain subject to substantial selection effects and calibration uncertainties (e.g. \citealt{Greene2024, Li2025}), they lie systematically above local black hole–galaxy scaling relations (e.g. \citealt{Pacucci2023, Harikane2023a}). Explaining such rapid early assembly lies beyond the reach of current large-volume cosmological simulations, which struggle to reconcile these high-redshift SMBHs with standard prescriptions for seeding, accretion, and feedback (e.g. \citealt{Ni2022, Gordon2025}). 

The first BHs are believed to have formed in the dense centres of very high redshift halos ($z \approx 20$--$30$) either as Population~III star remnants or via direct gas collapse (e.g.\ \citealt{Devecchi2012, Yue2014, Latif2016, Inayoshi2020, Regan2024}). Because both seeding channels depend sensitively on the gas reservoir at collapse, isocurvature perturbations will delay seed formation and suppress early accretion rates. Simulations that neglect these perturbations consequently over-predict early gas densities, accretion rates, and BH masses; incorporating isocurvature perturbations instead yields a more gradual, physically motivated build-up of the high-redshift SMBH population.

In summary, isocurvature perturbations modulate galaxy formation in $\Lambda$CDM by reducing the gas content available at collapse, thereby influencing any subsequent process that depends on the initial gas reservoir. The effects of isocurvature perturbations are strongest at early times and almost negligible today. Incorporating these perturbations introduces a physically motivated correction that is absent from almost all current cosmological simulations and may exacerbate existing high-redshift observational tensions. Given that these perturbations can be implemented straightforwardly using IC codes such as \texttt{MonofonIC}, we recommend their routine inclusion in future simulation suites to improve the fidelity of high-redshift galaxy formation predictions within $\Lambda$CDM.

\section*{Acknowledgements}
The authors would like to thank the referee for their careful reading of the manuscript and for their constructive comments, which significantly improved the clarity and presentation of this work. We would also like to thank Shaun Brown, Martin Rey, and Oliver Hahn for their helpful comments on an earlier version of this manuscript. We are also grateful to the \texttt{FLAMINGO} collaboration for enabling this work and for providing useful assistance throughout its development. OJ is supported by UK Science \& Technology Facilities Council (STFC) PhD Studentship ST/Y509346/1. ARJ and JCH are supported by STFC consolidated grant ST/X001075/1. AP is supported by the European Research Council (ERC) under the European Union’s Horizon 2020 research and innovation program (Grant agreement No. 818085 GMGalaxies). For the purpose of open access, the author has applied a Creative Commons Attribution (CC BY) licence to any accepted manuscript version arising from this submission. The simulations and analyses of this project made use of the DiRAC@Durham facility managed by the Institute for Computational Cosmology on behalf of the STFC DiRAC HPC Facility (www.dirac.ac.uk). The equipment was funded by BEIS capital funding via STFC capital grants ST/K00042X/1, ST/P002293/1, ST/R002371/1 and ST/S002502/1, Durham University and STFC operations grant ST/R000832/1. DiRAC is part of the National e-Infrastructure.

\section*{Data Availability}
The data underlying the figures in this manuscript are available from the corresponding author upon reasonable request. While the complete raw simulation dataset spans several terabytes and cannot be readily distributed, the \texttt{SOAP} halo catalogues amount to only a few tens of gigabytes.



\bibliographystyle{mnras}
\bibliography{refs} 

@article{Naoz2009,
  title = {Gas in simulations of high-redshift galaxies and minihaloes},
  volume = {399},
  ISSN = {1365-2966},
  url = {http://dx.doi.org/10.1111/j.1365-2966.2009.15282.x},
  DOI = {10.1111/j.1365-2966.2009.15282.x},
  number = {1},
  journal = {Monthly Notices of the Royal Astronomical Society},
  publisher = {Oxford University Press (OUP)},
  author = {Naoz,  Smadar and Barkana,  Rennan and Mesinger,  Andrei},
  year = {2009},
  month = oct,
  pages = {369–376}
}

@article{Gnedin1998,
  title = {Probing the Universe with the Ly  forest -- I. Hydrodynamics of the low-density intergalactic medium},
  volume = {296},
  ISSN = {1365-2966},
  url = {http://dx.doi.org/10.1046/j.1365-8711.1998.01249.x},
  DOI = {10.1046/j.1365-8711.1998.01249.x},
  number = {1},
  journal = {Monthly Notices of the Royal Astronomical Society},
  publisher = {Oxford University Press (OUP)},
  author = {Gnedin,  N. Y. and Hui,  L.},
  year = {1998},
  month = may,
  pages = {44–55}
}

@article{Gnedin2000,
   title={Effect of Reionization on Structure Formation in the Universe},
   volume={542},
   ISSN={1538-4357},
   url={http://dx.doi.org/10.1086/317042},
   DOI={10.1086/317042},
   number={2},
   journal={The Astrophysical Journal},
   publisher={American Astronomical Society},
   author={Gnedin, Nickolay Y.},
   year={2000},
   month=oct, pages={535–541} }

@article{Barkana2001,
   title={In the beginning: the first sources of light and the reionization of the universe},
   volume={349},
   ISSN={0370-1573},
   url={http://dx.doi.org/10.1016/S0370-1573(01)00019-9},
   DOI={10.1016/s0370-1573(01)00019-9},
   number={2},
   journal={Physics Reports},
   publisher={Elsevier BV},
   author={Barkana, Rennan and Loeb, Abraham},
   year={2001},
   month=jul, pages={125–238} }

@article{Lewis1999,
      author         = "Lewis, Antony and Challinor, Anthony and Lasenby,
                        Anthony",
      title          = "{Efficient computation of CMB anisotropies in closed FRW
                        models}",
      journal        = "\apj",
      volume         = "538",
      year           = "2000",
      pages          = "473-476",
      doi            = "10.1086/309179",
      eprint         = "astro-ph/9911177",
      archivePrefix  = "arXiv",
      primaryClass   = "astro-ph",
      SLACcitation   = "%%CITATION = ASTRO-PH/9911177;%%"
}

@article{Gupta2023,
   title={JWST early Universe observations and ΛCDM cosmology},
   volume={524},
   ISSN={1365-2966},
   url={http://dx.doi.org/10.1093/mnras/stad2032},
   DOI={10.1093/mnras/stad2032},
   number={3},
   journal={MNRAS},
   publisher={Oxford University Press (OUP)},
   author={Gupta, Rajendra P},
   year={2023},
   month=jul, pages={3385–3395} }

@article{Menci2022,
   title={High-redshift Galaxies from Early JWST Observations: Constraints on Dark Energy Models},
   volume={938},
   ISSN={2041-8213},
   url={http://dx.doi.org/10.3847/2041-8213/ac96e9},
   DOI={10.3847/2041-8213/ac96e9},
   number={1},
   journal={ApJ Letters},
   publisher={American Astronomical Society},
   author={Menci, N. and Castellano, M. and Santini, P. and Merlin, E. and Fontana, A. and Shankar, F.},
   year={2022},
   month=oct, pages={L5} }

@article{Lu2024,
      title={A comparison of pre-existing $\Lambda$CDM predictions with the abundance of {\it JWST} galaxies at high redshift}, 
      author={Shengdong Lu and Carlos S. Frenk and Sownak Bose and Cedric G. Lacey and Shaun Cole and Carlton M. Baugh and John C. Helly},
      year={2024},
      eprint={2406.02672},
      archivePrefix={arXiv},
      journal={arXiv},
      primaryClass={astro-ph.GA},
      url={https://arxiv.org/abs/2406.02672}, 
}

@article{Wang2023a,
      title={JWST high redshift galaxy observations have a strong tension with Planck CMB measurements}, 
      author={Deng Wang and Yizhou Liu},
      year={2023},
      eprint={2301.00347},
      archivePrefix={arXiv},
      primaryClass={astro-ph.CO},      
      journal={arXiv},
      url={https://arxiv.org/abs/2301.00347}, 
}

@article{Springel2005c,
   title={The cosmological simulation code gadget-2},
   volume={364},
   ISSN={1365-2966},
   url={http://dx.doi.org/10.1111/j.1365-2966.2005.09655.x},
   DOI={10.1111/j.1365-2966.2005.09655.x},
   number={4},
   journal={MNRAS},
   publisher={Oxford University Press (OUP)},
   author={Springel, Volker},
   year={2005},
   month=dec, pages={1105–1134} }

@article{MPGadget,
  doi = {10.5281/ZENODO.1451799},
  url = {https://zenodo.org/record/1451799},
  author = {Feng,  Yu and Bird,  Simeon and Anderson,  Lauren and Font-Ribera,  Andreu and Pedersen,  Chris},
  title = {MP-Gadget/MP-Gadget: A tag for getting a DOI},
  publisher = {Zenodo},
  journal={},
  year = {2018},
  copyright = {Open Access}
}

@article{Fialkov2012,
  title = {Impact of the relative motion between the dark matter and baryons on the first stars: semi-analytical modelling: Impact of relative motion on the first stars},
  volume = {424},
  ISSN = {0035-8711},
  url = {http://dx.doi.org/10.1111/j.1365-2966.2012.21318.x},
  DOI = {10.1111/j.1365-2966.2012.21318.x},
  number = {2},
  journal = {MNRAS},
  publisher = {Oxford University Press (OUP)},
  author = {Fialkov,  Anastasia and Barkana,  Rennan and Tseliakhovich,  Dmitriy and Hirata,  Christopher M.},
  year = {2012},
  month = jun,
  pages = {1335–1345}
}

@article{Park2020,
  title = {First Structure Formation under the Influence of Gas–Dark Matter Streaming Velocity and Density: Impact of the “Baryons Trace Dark Matter” Approximation},
  volume = {900},
  ISSN = {1538-4357},
  url = {http://dx.doi.org/10.3847/1538-4357/aba26e},
  DOI = {10.3847/1538-4357/aba26e},
  number = {1},
  journal = {ApJ},
  publisher = {American Astronomical Society},
  author = {Park, Hyunbae and Ahn, Kyungjin and Yoshida, Naoki and Hirano, Shingo},
  year = {2020},
  month = aug,
  pages = {30}
}

@article{Spergel2003,
  title = {First‐Year Wilkinson Microwave Anisotropy Probe (WMAP) Observations: Determination of Cosmological Parameters},
  volume = {148},
  ISSN = {1538-4365},
  url = {http://dx.doi.org/10.1086/377226},
  DOI = {10.1086/377226},
  number = {1},
  journal = {ApJ Supplement Series},
  publisher = {American Astronomical Society},
  author = {Spergel,  D. N. and Verde,  L. and Peiris,  H. V. and Komatsu,  E. and Nolta,  M. R. and Bennett,  C. L. and Halpern,  M. and Hinshaw,  G. and Jarosik,  N. and Kogut,  A. and Limon,  M. and Meyer,  S. S. and Page,  L. and Tucker,  G. S. and Weiland,  J. L. and Wollack,  E. and Wright,  E. L.},
  year = {2003},
  month = sep,
  pages = {175–194}
}

@article{Munoz2019,
  title = {Robust velocity-induced acoustic oscillations at cosmic dawn},
  volume = {100},
  ISSN = {2470-0029},
  url = {http://dx.doi.org/10.1103/PhysRevD.100.063538},
  DOI = {10.1103/physrevd.100.063538},
  number = {6},
  journal = {Physical Review D},
  publisher = {American Physical Society (APS)},
  author = {Muñoz,  Julian B.},
  year = {2019},
  month = sep 
}

@article{Schauer2023,
   title={Dwarf Galaxy Formation with and without Dark Matter–Baryon Streaming Velocities},
   volume={950},
   ISSN={1538-4357},
   url={http://dx.doi.org/10.3847/1538-4357/accc2c},
   DOI={10.3847/1538-4357/accc2c},
   number={1},
   journal={ApJ},
   publisher={American Astronomical Society},
   author={Schauer, Anna T. P. and Boylan-Kolchin, Michael and Colston, Katelyn and Sameie, Omid and Bromm, Volker and Bullock, James S. and Wetzel, Andrew},
   year={2023},
   month=jun, pages={20} }

@article{Blazek2016,
   title={Streaming Velocities and the Baryon Acoustic Oscillation Scale},
   volume={116},
   ISSN={1079-7114},
   url={http://dx.doi.org/10.1103/PhysRevLett.116.121303},
   DOI={10.1103/physrevlett.116.121303},
   number={12},
   journal={Physical Review Letters},
   publisher={American Physical Society (APS)},
   author={Blazek, Jonathan A. and McEwen, Joseph E. and Hirata, Christopher M.},
   year={2016},
   month=mar }

@article{Schmidt2016,
   title={Effect of relative velocity and density perturbations between baryons and dark matter on the clustering of galaxies},
   volume={94},
   ISSN={2470-0029},
   url={http://dx.doi.org/10.1103/PhysRevD.94.063508},
   DOI={10.1103/physrevd.94.063508},
   number={6},
   journal={Physical Review D},
   publisher={American Physical Society (APS)},
   author={Schmidt, Fabian},
   year={2016},
   month=sep }

@article{Heinrich2019,
   title={BAO modulation as a probe of compensated isocurvature perturbations},
   volume={100},
   ISSN={2470-0029},
   url={http://dx.doi.org/10.1103/PhysRevD.100.063503},
   DOI={10.1103/physrevd.100.063503},
   number={6},
   journal={Physical Review D},
   publisher={American Physical Society (APS)},
   author={Heinrich, Chen and Schmittfull, Marcel},
   year={2019},
   month=sep }

@article{Kannan2021,
   title={Introducing the <scp>thesan</scp> project: radiation-magnetohydrodynamic simulations of the epoch of reionization},
   volume={511},
   ISSN={1365-2966},
   url={http://dx.doi.org/10.1093/mnras/stab3710},
   DOI={10.1093/mnras/stab3710},
   number={3},
   journal={MNRAS},
   publisher={Oxford University Press (OUP)},
   author={Kannan, R and Garaldi, E and Smith, A and Pakmor, R and Springel, V and Vogelsberger, M and Hernquist, L},
   year={2021},
   month=dec, pages={4005–4030} }

@article{Lovell2020,
   title={First Light And Reionization Epoch Simulations (FLARES) – I. Environmental dependence of high-redshift galaxy evolution},
   volume={500},
   ISSN={1365-2966},
   url={http://dx.doi.org/10.1093/mnras/staa3360},
   DOI={10.1093/mnras/staa3360},
   number={2},
   journal={MNRAS},
   publisher={Oxford University Press (OUP)},
   author={Lovell, Christopher C and Vijayan, Aswin P and Thomas, Peter A and Wilkins, Stephen M and Barnes, David J and Irodotou, Dimitrios and Roper, Will},
   year={2020},
   month=oct, pages={2127–2145} }

@article{Kautz2024,
   title={Phasing the Giant Magellan Telescope: lab experiments and first on-sky demonstration},
   volume={10},
   ISSN={2329-4124},
   url={http://dx.doi.org/10.1117/1.JATIS.10.4.049005},
   DOI={10.1117/1.jatis.10.4.049005},
   number={04},
   journal={Journal of Astronomical Telescopes, Instruments, and Systems},
   publisher={SPIE-Intl Soc Optical Eng},
   author={Kautz, Maggie Y. and Haffert, Sebastiaan Y. and Close, Laird M. and Males, Jared R. and Guyon, Olivier and Hedglen, Alexander D. and Gasho, Victor and Demers, Richard and Bouchez, Antonin and Quirós-Pacheco, Fernando and Plantet, Cédric and McLeod, Avalon L. and Kueny, Jay K. and Li, Jialin and Liberman, Joshua and Long, Joseph D. and Lumbres, Jennifer and McEwen, Eden A. and Pearce, Logan A. and Schatz, Lauren and Schurter, Patricio and Sitarski, Breann and Twitchell, Katie and Van Gorkom, Kyle},
   year={2024},
   month=nov }

@article{Padovani2023,
   title={The Extremely Large Telescope},
   volume={64},
   ISSN={1366-5812},
   url={http://dx.doi.org/10.1080/00107514.2023.2266921},
   DOI={10.1080/00107514.2023.2266921},
   number={1},
   journal={Contemporary Physics},
   publisher={Informa UK Limited},
   author={Padovani, Paolo and Cirasuolo, Michele},
   year={2023},
   month=jan, pages={47–64}}

@article{Valkenburg2017,
   title={Accurate initial conditions in mixed dark matter–baryon simulations},
   volume={467},
   ISSN={1365-2966},
   url={http://dx.doi.org/10.1093/mnras/stx376},
   DOI={10.1093/mnras/stx376},
   number={4},
   journal={MNRAS},
   publisher={Oxford University Press (OUP)},
   author={Valkenburg, Wessel and Villaescusa-Navarro, Francisco},
   year={2017},
   month=feb, pages={4401–4409} }

@article{Smith2017,
   title={Baryons still trace dark matter: Probing CMB lensing maps for hidden isocurvature},
   volume={96},
   ISSN={2470-0029},
   url={http://dx.doi.org/10.1103/PhysRevD.96.083508},
   DOI={10.1103/physrevd.96.083508},
   number={8},
   journal={Physical Review D},
   publisher={American Physical Society (APS)},
   author={Smith, Tristan L. and Muñoz, Julian B. and Smith, Rhiannon and Yee, Kyle and Grin, Daniel},
   year={2017},
   month=oct }

@article{Chen2019,
   title={Biased tracers of two fluids in the Lagrangian picture},
   volume={2019},
   ISSN={1475-7516},
   url={http://dx.doi.org/10.1088/1475-7516/2019/06/006},
   DOI={10.1088/1475-7516/2019/06/006},
   number={06},
   journal={JCAP},
   publisher={IOP Publishing},
   author={Chen, Shi-Fan and Castorina, Emanuele and White, Martin},
   year={2019},
   month=jun, pages={006–006} }

@article{Fialkov2014,
   title={Supersonic relative velocity between dark matter and baryons: A review},
   volume={23},
   ISSN={1793-6594},
   url={http://dx.doi.org/10.1142/S0218271814300171},
   DOI={10.1142/s0218271814300171},
   number={08},
   journal={International Journal of Modern Physics D},
   publisher={World Scientific Pub Co Pte Ltd},
   author={Fialkov, Anastasia},
   year={2014},
   month=jul, pages={1430017} }

@article{Schaye2010,
   title={The physics driving the cosmic star formation history},
   volume={402},
   ISSN={1365-2966},
   url={http://dx.doi.org/10.1111/j.1365-2966.2009.16029.x},
   DOI={10.1111/j.1365-2966.2009.16029.x},
   number={3},
   journal={MNRAS},
   publisher={Oxford University Press (OUP)},
   author={Schaye, Joop and Vecchia, Claudio Dalla and Booth, C. M. and Wiersma, Robert P. C. and Theuns, Tom and Haas, Marcel R. and Bertone, Serena and Duffy, Alan R. and McCarthy, I. G. and van de Voort, Freeke},
   year={2010},
   month=mar, pages={1536–1560} }

@article{Madau2014,
   title={Cosmic Star-Formation History},
   volume={52},
   ISSN={1545-4282},
   url={http://dx.doi.org/10.1146/annurev-astro-081811-125615},
   DOI={10.1146/annurev-astro-081811-125615},
   number={1},
   journal={ARA&A},
   publisher={Annual Reviews},
   author={Madau, Piero and Dickinson, Mark},
   year={2014},
   month=aug, pages={415–486} }

@book{Weinberg2008,
  author    = {Weinberg, Steven},
  title     = {Cosmology},
  year      = {2008},
  publisher = {Oxford University Press},
  address   = {Oxford},
  isbn      = {9780198526827},
}

@article{Planck2020,
   title={Planck2018 results: I. Overview and the cosmological legacy ofPlanck},
   volume={641},
   ISSN={1432-0746},
   url={http://dx.doi.org/10.1051/0004-6361/201833880},
   DOI={10.1051/0004-6361/201833880},
   journal={A&A},
   publisher={EDP Sciences},
   author={Aghanim, N. and Akrami, Y. and Arroja, F. and Ashdown, M. and Aumont, J. and Baccigalupi, C. and Ballardini, M. and Banday, A. J. and Barreiro, R. B. and Bartolo, N. and Basak, S. and Battye, R. and Benabed, K. and Bernard, J.-P. and Bersanelli, M. and Bielewicz, P. and Bock, J. J. and Bond, J. R. and Borrill, J. and Bouchet, F. R. and Boulanger, F. and Bucher, M. and Burigana, C. and Butler, R. C. and Calabrese, E. and Cardoso, J.-F. and Carron, J. and Casaponsa, B. and Challinor, A. and Chiang, H. C. and Colombo, L. P. L. and Combet, C. and Contreras, D. and Crill, B. P. and Cuttaia, F. and de Bernardis, P. and de Zotti, G. and Delabrouille, J. and Delouis, J.-M. and Désert, F.-X. and Di Valentino, E. and Dickinson, C. and Diego, J. M. and Donzelli, S. and Doré, O. and Douspis, M. and Ducout, A. and Dupac, X. and Efstathiou, G. and Elsner, F. and Enßlin, T. A. and Eriksen, H. K. and Falgarone, E. and Fantaye, Y. and Fergusson, J. and Fernandez-Cobos, R. and Finelli, F. and Forastieri, F. and Frailis, M. and Franceschi, E. and Frolov, A. and Galeotta, S. and Galli, S. and Ganga, K. and Génova-Santos, R. T. and Gerbino, M. and Ghosh, T. and González-Nuevo, J. and Górski, K. M. and Gratton, S. and Gruppuso, A. and Gudmundsson, J. E. and Hamann, J. and Handley, W. and Hansen, F. K. and Helou, G. and Herranz, D. and Hildebrandt, S. R. and Hivon, E. and Huang, Z. and Jaffe, A. H. and Jones, W. C. and Karakci, A. and Keihänen, E. and Keskitalo, R. and Kiiveri, K. and Kim, J. and Kisner, T. S. and Knox, L. and Krachmalnicoff, N. and Kunz, M. and Kurki-Suonio, H. and Lagache, G. and Lamarre, J.-M. and Langer, M. and Lasenby, A. and Lattanzi, M. and Lawrence, C. R. and Le Jeune, M. and Leahy, J. P. and Lesgourgues, J. and Levrier, F. and Lewis, A. and Liguori, M. and Lilje, P. B. and Lilley, M. and Lindholm, V. and López-Caniego, M. and Lubin, P. M. and Ma, Y.-Z. and Macías-Pérez, J. F. and Maggio, G. and Maino, D. and Mandolesi, N. and Mangilli, A. and Marcos-Caballero, A. and Maris, M. and Martin, P. G. and Martinelli, M. and Martínez-González, E. and Matarrese, S. and Mauri, N. and McEwen, J. D. and Meerburg, P. D. and Meinhold, P. R. and Melchiorri, A. and Mennella, A. and Migliaccio, M. and Millea, M. and Mitra, S. and Miville-Deschênes, M.-A. and Molinari, D. and Moneti, A. and Montier, L. and Morgante, G. and Moss, A. and Mottet, S. and Münchmeyer, M. and Natoli, P. and Nørgaard-Nielsen, H. U. and Oxborrow, C. A. and Pagano, L. and Paoletti, D. and Partridge, B. and Patanchon, G. and Pearson, T. J. and Peel, M. and Peiris, H. V. and Perrotta, F. and Pettorino, V. and Piacentini, F. and Polastri, L. and Polenta, G. and Puget, J.-L. and Rachen, J. P. and Reinecke, M. and Remazeilles, M. and Renault, C. and Renzi, A. and Rocha, G. and Rosset, C. and Roudier, G. and Rubiño-Martín, J. A. and Ruiz-Granados, B. and Salvati, L. and Sandri, M. and Savelainen, M. and Scott, D. and Shellard, E. P. S. and Shiraishi, M. and Sirignano, C. and Sirri, G. and Spencer, L. D. and Sunyaev, R. and Suur-Uski, A.-S. and Tauber, J. A. and Tavagnacco, D. and Tenti, M. and Terenzi, L. and Toffolatti, L. and Tomasi, M. and Trombetti, T. and Valiviita, J. and Van Tent, B. and Vibert, L. and Vielva, P. and Villa, F. and Vittorio, N. and Wandelt, B. D. and Wehus, I. K. and White, M. and White, S. D. M. and Zacchei, A. and Zonca, A.},
   year={2020},
   month=sep, pages={A1} }

@article{Bahe2022,
   title={The importance of black hole repositioning for galaxy formation simulations},
   volume={516},
   ISSN={1365-2966},
   url={http://dx.doi.org/10.1093/mnras/stac1339},
   DOI={10.1093/mnras/stac1339},
   number={1},
   journal={MNRAS},
   publisher={Oxford University Press (OUP)},
   author={Bahé, Yannick M and Schaye, Joop and Schaller, Matthieu and Bower, Richard G and Borrow, Josh and Chaikin, Evgenii and Kugel, Roi and Nobels, Folkert and Ploeckinger, Sylvia},
   year={2022},
   month=may, pages={167–184} }

@article{Abbott2022,
  title = {Dark Energy Survey Year 3 results: Cosmological constraints from galaxy clustering and weak lensing},
  volume = {105},
  ISSN = {2470-0029},
  url = {http://dx.doi.org/10.1103/PhysRevD.105.023520},
  DOI = {10.1103/physrevd.105.023520},
  number = {2},
  journal = {Physical Review D},
  publisher = {American Physical Society (APS)},
  author = {Abbott,  T.M.C. and Aguena,  M. and Alarcon,  A. and Allam},
  year = {2022},
  month = jan 
}

@article{Schaye2025,
      title={The COLIBRE project: cosmological hydrodynamical simulations of galaxy formation and evolution}, 
      author={Joop Schaye and Evgenii Chaikin and Matthieu Schaller and Sylvia Ploeckinger and Filip Huško and Rob McGibbon and James W. Trayford and Alejandro Benítez-Llambay and Camila Correa and Carlos S. Frenk and Alexander J. Richings and Victor J. Forouhar Moreno and Yannick M. Bahé and Josh Borrow and Anna Durrant and Andrea Gebek and John C. Helly and Adrian Jenkins and Cedric G. Lacey and Aaron Ludlow and Folkert S. J. Nobels},
      year={2025},
      journal={arXiv preprints},
      eprint={2508.21126},
      archivePrefix={arXiv},
      pages={arXiv:2508.21126},
      primaryClass={astro-ph.GA},
      url={https://arxiv.org/abs/2508.21126}, 
}

@article{Chaikin2025a,
      title={COLIBRE: calibrating subgrid feedback in cosmological simulations that include a cold gas phase}, 
      author={Evgenii Chaikin and Joop Schaye and Matthieu Schaller and Sylvia Ploeckinger and Yannick M. Bahé and Alejandro Benítez-Llambay and Camila Correa and Victor J. Forouhar Moreno and Carlos S. Frenk and Filip Huško and Roi Kugel and Robert McGibbon and Alexander J. Richings and James W. Trayford and Josh Borrow and Robert A. Crain and John C. Helly and Cedric G. Lacey and Aaron Ludlow and Folkert S. J. Nobels},
      year={2025},
      eprint={2509.04067},
      archivePrefix={arXiv},
      journal={arXiv preprints},
      pages={arXiv:2509.04067},
      primaryClass={astro-ph.GA},
      url={https://arxiv.org/abs/2509.04067}, 
}

@article{Barreira2020b,
   title={Compensated isocurvature perturbations in the galaxy power spectrum},
   volume={2020},
   ISSN={1475-7516},
   url={http://dx.doi.org/10.1088/1475-7516/2020/07/049},
   DOI={10.1088/1475-7516/2020/07/049},
   number={07},
   journal={JCAP},
   publisher={IOP Publishing},
   author={Barreira, Alexandre and Cabass, Giovanni and Lozanov, Kaloian D. and Schmidt, Fabian},
   year={2020},
   month=jul, pages={049–049} }

@article{Weinberger2025,
   title={Accretion onto supermassive and intermediate-mass black holes in cosmological simulations},
   volume={700},
   ISSN={1432-0746},
   url={http://dx.doi.org/10.1051/0004-6361/202554174},
   DOI={10.1051/0004-6361/202554174},
   journal={A&A},
   publisher={EDP Sciences},
   author={Weinberger, R. and Bhowmick, A. and Blecha, L. and Bryan, G. and Buchner, J. and Hernquist, L. and Hlavacek-Larrondo, J. and Springel, V.},
   year={2025},
   month=aug, pages={A52} }

@article{Pallottini2022,
   title={A survey of high-z galaxies: SERRA simulations},
   ISSN={1365-2966},
   url={http://dx.doi.org/10.1093/mnras/stac1281},
   DOI={10.1093/mnras/stac1281},
   journal={MNRAS},
   publisher={Oxford University Press (OUP)},
   author={Pallottini, A and Ferrara, A and Gallerani, S and Behrens, C and Kohandel, M and Carniani, S and Vallini, L and Salvadori, S and Gelli, V and Sommovigo, L and D’Odorico, V and Di Mascia, F and Pizzati, E},
   year={2022},
   month=may }

@article{Barreira2020,
   title={Baryon-CDM isocurvature galaxy bias with IllustrisTNG},
   volume={2020},
   ISSN={1475-7516},
   url={http://dx.doi.org/10.1088/1475-7516/2020/02/005},
   DOI={10.1088/1475-7516/2020/02/005},
   number={02},
   journal={JCAP},
   publisher={IOP Publishing},
   author={Barreira, Alexandre and Cabass, Giovanni and Nelson, Dylan and Schmidt, Fabian},
   year={2020},
   month=feb, pages={005–005} }

@article{Behroozi2013,
   title={THE AVERAGE STAR FORMATION HISTORIES OF GALAXIES IN DARK MATTER HALOS FROMz= 0-8},
   volume={770},
   ISSN={1538-4357},
   url={http://dx.doi.org/10.1088/0004-637X/770/1/57},
   DOI={10.1088/0004-637x/770/1/57},
   number={1},
   journal={ApJ},
   publisher={American Astronomical Society},
   author={Behroozi, Peter S. and Wechsler, Risa H. and Conroy, Charlie},
   year={2013},
   month=may, pages={57} }

@article{Davis1985,
  title = {The evolution of large-scale structure in a universe dominated by cold dark matter},
  volume = {292},
  ISSN = {1538-4357},
  url = {http://dx.doi.org/10.1086/163168},
  DOI = {10.1086/163168},
  journal = {ApJ},
  publisher = {American Astronomical Society},
  author = {Davis,  M. and Efstathiou,  G. and Frenk,  C. S. and White,  S. D. M.},
  year = {1985},
  month = may,
  pages = {371}
}

@article{Navarro1997,
   title={A Universal Density Profile from Hierarchical Clustering},
   volume={490},
   ISSN={1538-4357},
   url={http://dx.doi.org/10.1086/304888},
   DOI={10.1086/304888},
   number={2},
   journal={ApJ},
   publisher={American Astronomical Society},
   author={Navarro, Julio F. and Frenk, Carlos S. and White, Simon D. M.},
   year={1997},
   month=dec, pages={493–508} }

@article{Stark2025,
      title={Observations of the First Galaxies in the Era of JWST}, 
      author={Daniel P. Stark and Michael W. Topping and Ryan Endsley and Mengtao Tang},
      year={2025},
      eprint={2501.17078},
      archivePrefix={arXiv},
      journal={arXiv},
      primaryClass={astro-ph.GA},
      url={https://arxiv.org/abs/2501.17078}, 
}

@article{Bardeen1986,
  title = {The statistics of peaks of Gaussian random fields},
  volume = {304},
  ISSN = {1538-4357},
  url = {http://dx.doi.org/10.1086/164143},
  DOI = {10.1086/164143},
  journal = {ApJ},
  publisher = {American Astronomical Society},
  author = {Bardeen,  J. M. and Bond,  J. R. and Kaiser,  N. and Szalay,  A. S.},
  year = {1986},
  month = may,
  pages = {15}
}

@article{Bond1984,
  title = {Cosmic background radiation anisotropies in universes dominated by nonbaryonic dark matter},
  volume = {285},
  ISSN = {1538-4357},
  url = {http://dx.doi.org/10.1086/184362},
  DOI = {10.1086/184362},
  journal = {ApJ},
  publisher = {American Astronomical Society},
  author = {Bond,  J. R. and Efstathiou,  G.},
  year = {1984},
  month = oct,
  pages = {L45}
}

@article{Peebles1970,
  title = {Primeval Adiabatic Perturbation in an Expanding Universe},
  volume = {162},
  ISSN = {1538-4357},
  url = {http://dx.doi.org/10.1086/150713},
  DOI = {10.1086/150713},
  journal = {ApJ},
  publisher = {American Astronomical Society},
  author = {Peebles,  P. J. E. and Yu,  J. T.},
  year = {1970},
  month = dec,
  pages = {815}
}

@article{Harikane2023a,
  title = {A Comprehensive Study of Galaxies at z ∼ 9–16 Found in the Early JWST Data: Ultraviolet Luminosity Functions and Cosmic Star Formation History at the Pre-reionization Epoch},
  volume = {265},
  ISSN = {1538-4365},
  url = {http://dx.doi.org/10.3847/1538-4365/acaaa9},
  DOI = {10.3847/1538-4365/acaaa9},
  number = {1},
  journal = {ApJ Supplement Series},
  publisher = {American Astronomical Society},
  author = {Harikane,  Yuichi and Ouchi,  Masami and Oguri,  Masamune and Ono,  Yoshiaki and Nakajima,  Kimihiko and Isobe,  Yuki and Umeda,  Hiroya and Mawatari,  Ken and Zhang,  Yechi},
  year = {2023},
  month = feb,
  pages = {5}
}

@article{Springel2005b,
   title={Modelling feedback from stars and black holes in galaxy mergers},
   volume={361},
   ISSN={1365-2966},
   url={http://dx.doi.org/10.1111/j.1365-2966.2005.09238.x},
   DOI={10.1111/j.1365-2966.2005.09238.x},
   number={3},
   journal={MNRAS},
   publisher={Oxford University Press (OUP)},
   author={Springel, Volker and Di Matteo, Tiziana and Hernquist, Lars},
   year={2005},
   month=aug, pages={776–794} }

@article{Finkelstein2023,
  title = {CEERS Key Paper. I. An Early Look into the First 500 Myr of Galaxy Formation with JWST},
  volume = {946},
  ISSN = {2041-8213},
  url = {http://dx.doi.org/10.3847/2041-8213/acade4},
  DOI = {10.3847/2041-8213/acade4},
  number = {1},
  journal = {ApJ Letters},
  publisher = {American Astronomical Society},
  author = {Finkelstein,  Steven L. and Bagley,  Micaela B. and Ferguson,  Henry C. and Wilkins,  Stephen M. and Kartaltepe,  Jeyhan S. and Papovich,  Casey and Yung,  L. Y. Aaron and Arrabal Haro,  Pablo and Behroozi,  Peter and Dickinson,  Mark and Kocevski,  Dale D. and Koekemoer,  Anton M. and Larson,  Rebecca L. and Le Bail,  Aurélien and Morales,  Alexa M. and Pérez-González,  Pablo G. and Burgarella,  Denis and Davé,  Romeel and Hirschmann,  Michaela and Somerville,  Rachel S. and Wuyts,  Stijn and Bromm,  Volker and Casey,  Caitlin M. and Fontana,  Adriano and Fujimoto,  Seiji and Gardner,  Jonathan P. and Giavalisco,  Mauro and Grazian,  Andrea and Grogin,  Norman A. and Hathi,  Nimish P. and Hutchison,  Taylor A. and Jha,  Saurabh W. and Jogee,  Shardha and Kewley,  Lisa J. and Kirkpatrick,  Allison and Long,  Arianna S. and Lotz,  Jennifer M. and Pentericci,  Laura and Pierel,  Justin D. R. and Pirzkal,  Nor and Ravindranath,  Swara and Ryan,  Russell E. and Trump,  Jonathan R. and Yang,  Guang and Bhatawdekar,  Rachana and Bisigello,  Laura and Buat,  Véronique and Calabrò,  Antonello and Castellano,  Marco and Cleri,  Nikko J. and Cooper,  M. C. and Croton,  Darren and Daddi,  Emanuele and Dekel,  Avishai and Elbaz,  David and Franco,  Maximilien and Gawiser,  Eric and Holwerda,  Benne W. and Huertas-Company,  Marc and Jaskot,  Anne E. and Leung,  Gene C. K. and Lucas,  Ray A. and Mobasher,  Bahram and Pandya,  Viraj and Tacchella,  Sandro and Weiner,  Benjamin J. and Zavala,  Jorge A.},
  year = {2023},
  month = mar,
  pages = {L13}
}

@article{Bunker2019, 
    title={Spectroscopy with the JWST Advanced Deep Extragalactic Survey (JADES) - the NIRSpec/NIRCAM GTO galaxy evolution project}, 
    volume={15}, 
    DOI={10.1017/S1743921319009463}, 
    number={S352}, 
    journal={Proceedings of the International Astronomical Union},
    author={Bunker, Andrew J.}, 
    year={2019}, 
    pages={342–346}
}

@article{McCaffrey2023,
   title={No Tension: JWST Galaxies at z&gt;10 Consistent with Cosmological Simulations},
   volume={6},
   ISSN={2565-6120},
   url={http://dx.doi.org/10.21105/astro.2304.13755},
   DOI={10.21105/astro.2304.13755},
   journal={The Open Journal of Astrophysics},
   publisher={Maynooth University},
   author={McCaffrey, Joe and Hardin, Samantha and Wise, John H. and Regan, John A.},
   year={2023},
   month=sep }

@article{Kennicutt1998,
   title={The Global Schmidt Law in Star‐forming Galaxies},
   volume={498},
   ISSN={1538-4357},
   url={http://dx.doi.org/10.1086/305588},
   DOI={10.1086/305588},
   number={2},
   journal={ApJ},
   publisher={American Astronomical Society},
   author={Kennicutt, Jr., Robert C.},
   year={1998},
   month=may, pages={541–552} }

@article{JWST2023,
   title={The James Webb Space Telescope Mission},
   volume={135},
   ISSN={1538-3873},
   url={http://dx.doi.org/10.1088/1538-3873/acd1b5},
   DOI={10.1088/1538-3873/acd1b5},
   number={1048},
   journal={PASP},
   publisher={IOP Publishing},
   author={Gardner, Jonathan P. and Mather, John C. and Abbott, Randy and Abell, James S. and Abernathy, Mark and Abney, et al.},
   year={2023},
   month=jun, pages={068001} }

@article{Behroozi2019,
   title={UniverseMachine: The correlation between galaxy growth and dark matter halo assembly from z = 0−10},
   volume={488},
   ISSN={1365-2966},
   url={http://dx.doi.org/10.1093/mnras/stz1182},
   DOI={10.1093/mnras/stz1182},
   number={3},
   journal={MNRAS},
   publisher={Oxford University Press (OUP)},
   author={Behroozi, Peter and Wechsler, Risa H and Hearin, Andrew P and Conroy, Charlie},
   year={2019},
   month=may, pages={3143–3194} }

@book{Coles2002,
  title={Cosmology: The Origin and Evolution of Cosmic Structure},
  author={Coles, Peter and Lucchin, Francesco},
  year={2002},
  publisher={Wiley}
}

@article{Fakhouri2010,
   title={The merger rates and mass assembly histories of dark matter haloes in the two Millennium simulations: Merger rates},
   volume={406},
   ISSN={0035-8711},
   url={http://dx.doi.org/10.1111/j.1365-2966.2010.16859.x},
   DOI={10.1111/j.1365-2966.2010.16859.x},
   number={4},
   journal={MNRAS},
   publisher={Oxford University Press (OUP)},
   author={Fakhouri, Onsi and Ma, Chung-Pei and Boylan-Kolchin, Michael},
   year={2010},
   month=jun, pages={2267–2278} }

@article{Eisenstein2005,
   title={Detection of the Baryon Acoustic Peak in the Large‐Scale Correlation Function of SDSS Luminous Red Galaxies},
   volume={633},
   ISSN={1538-4357},
   url={http://dx.doi.org/10.1086/466512},
   DOI={10.1086/466512},
   number={2},
   journal={ApJ},
   publisher={American Astronomical Society},
   author={Eisenstein, Daniel J. and Zehavi, Idit and Hogg, David W. and Scoccimarro, Roman and Blanton, Michael R. and Nichol, Robert C. and Scranton, Ryan and Seo, Hee‐Jong and Tegmark, Max and Zheng, Zheng and Anderson, Scott F. and Annis, Jim and Bahcall, Neta and Brinkmann, Jon and Burles, Scott and Castander, Francisco J. and Connolly, Andrew and Csabai, Istvan and Doi, Mamoru and Fukugita, Masataka and Frieman, Joshua A. and Glazebrook, Karl and Gunn, James E. and Hendry, John S. and Hennessy, Gregory and Ivezić, Zeljko and Kent, Stephen and Knapp, Gillian R. and Lin, Huan and Loh, Yeong‐Shang and Lupton, Robert H. and Margon, Bruce and McKay, Timothy A. and Meiksin, Avery and Munn, Jeffery A. and Pope, Adrian and Richmond, Michael W. and Schlegel, David and Schneider, Donald P. and Shimasaku, Kazuhiro and Stoughton, Christopher and Strauss, Michael A. and SubbaRao, Mark and Szalay, Alexander S. and Szapudi, Istvan and Tucker, Douglas L. and Yanny, Brian and York, Donald G.},
   year={2005},
   month=nov, pages={560–574} }

@article{Cole2005,
   title={The 2dF Galaxy Redshift Survey: power-spectrum analysis of the final data set and cosmological implications},
   volume={362},
   ISSN={1365-2966},
   url={http://dx.doi.org/10.1111/j.1365-2966.2005.09318.x},
   DOI={10.1111/j.1365-2966.2005.09318.x},
   number={2},
   journal={MNRAS},
   publisher={Oxford University Press (OUP)},
   author={Cole, Shaun and Percival, Will J. and Peacock, John A. and Norberg, Peder and Baugh, Carlton M. and Frenk, Carlos S. and Baldry, Ivan and Bland-Hawthorn, Joss and Bridges, Terry and Cannon, Russell and Colless, Matthew and Collins, Chris and Couch, Warrick and Cross, Nicholas J. G. and Dalton, Gavin and Eke, Vincent R. and De Propris, Roberto and Driver, Simon P. and Efstathiou, George and Ellis, Richard S. and Glazebrook, Karl and Jackson, Carole and Jenkins, Adrian and Lahav, Ofer and Lewis, Ian and Lumsden, Stuart and Maddox, Steve and Madgwick, Darren and Peterson, Bruce A. and Sutherland, Will and Taylor, Keith},
   year={2005},
   month=sep, pages={505–534} }

@article{Eisenstein1998,
   title={Baryonic Features in the Matter Transfer Function},
   volume={496},
   ISSN={1538-4357},
   url={http://dx.doi.org/10.1086/305424},
   DOI={10.1086/305424},
   number={2},
   journal={ApJ},
   publisher={American Astronomical Society},
   author={Eisenstein, Daniel J. and Hu, Wayne},
   year={1998},
   month=apr, pages={605–614} }

@article{McCarthy2017,
   title={The bahamas project: calibrated hydrodynamical simulations for large-scale structure cosmology},
   volume={465},
   ISSN={1365-2966},
   url={http://dx.doi.org/10.1093/mnras/stw2792},
   DOI={10.1093/mnras/stw2792},
   number={3},
   journal={MNRAS},
   publisher={Oxford University Press (OUP)},
   author={McCarthy, Ian G. and Schaye, Joop and Bird, Simeon and Le Brun, Amandine M. C.},
   year={2017},
   month=oct, pages={2936–2965} }

@article{Khoraminezhad2021,
   title={Quantifying the impact of baryon-CDM perturbations on halo clustering and baryon fraction},
   volume={2021},
   ISSN={1475-7516},
   url={http://dx.doi.org/10.1088/1475-7516/2021/03/023},
   DOI={10.1088/1475-7516/2021/03/023},
   number={03},
   journal={JCAP},
   publisher={IOP Publishing},
   author={Khoraminezhad, Hasti and Lazeyras, Titouan and Angulo, Raul E. and Hahn, Oliver and Viel, Matteo},
   year={2021},
   month=mar, pages={023} }

@article{Gunn1972,
  title = {On the Infall of Matter Into Clusters of Galaxies and Some Effects on Their Evolution},
  volume = {176},
  ISSN = {1538-4357},
  url = {http://dx.doi.org/10.1086/151605},
  DOI = {10.1086/151605},
  journal = {ApJ},
  publisher = {American Astronomical Society},
  author = {Gunn,  James E. and Gott,  J. Richard,  III},
  year = {1972},
  month = aug,
  pages = {1}
}

@article{Pakmor2023,
   title={The MillenniumTNG Project: the hydrodynamical full physics simulation and a first look at its galaxy clusters},
   volume={524},
   ISSN={1365-2966},
   url={http://dx.doi.org/10.1093/mnras/stac3620},
   DOI={10.1093/mnras/stac3620},
   number={2},
   journal={MNRAS},
   publisher={Oxford University Press (OUP)},
   author={Pakmor, Rüdiger and Springel, Volker and Coles, Jonathan P and Guillet, Thomas and Pfrommer, Christoph and Bose, Sownak and Barrera, Monica and Delgado, Ana Maria and Ferlito, Fulvio and Frenk, Carlos and Hadzhiyska, Boryana and Hernández-Aguayo, César and Hernquist, Lars and Kannan, Rahul and White, Simon D M},
   year={2023},
   month=jul, pages={2539–2555} }

@article{Pillepich2017,
   title={Simulating galaxy formation with the IllustrisTNG model},
   volume={473},
   ISSN={1365-2966},
   url={http://dx.doi.org/10.1093/mnras/stx2656},
   DOI={10.1093/mnras/stx2656},
   number={3},
   journal={MNRAS},
   publisher={Oxford University Press (OUP)},
   author={Pillepich, Annalisa and Springel, Volker and Nelson, Dylan and Genel, Shy and Naiman, Jill and Pakmor, Rüdiger and Hernquist, Lars and Torrey, Paul and Vogelsberger, Mark and Weinberger, Rainer and Marinacci, Federico},
   year={2017},
   month=oct, pages={4077–4106} }

@article{Tseliakhovich2010,
   title={Relative velocity of dark matter and baryonic fluids and the formation of the first structures},
   volume={82},
   ISSN={1550-2368},
   url={http://dx.doi.org/10.1103/PhysRevD.82.083520},
   DOI={10.1103/physrevd.82.083520},
   number={8},
   journal={Physical Review D},
   publisher={American Physical Society (APS)},
   author={Tseliakhovich, Dmitriy and Hirata, Christopher},
   year={2010},
   month=oct }

@article{Angulo2013,
   title={How closely do baryons follow dark matter on large scales?},
   volume={434},
   ISSN={1365-2966},
   url={http://dx.doi.org/10.1093/mnras/stt1135},
   DOI={10.1093/mnras/stt1135},
   number={2},
   journal={MNRAS},
   publisher={Oxford University Press (OUP)},
   author={Angulo, Raul E. and Hahn, Oliver and Abel, Tom},
   year={2013},
   month=jul, pages={1756–1764} }

@article{Michaux2020,
   title={Accurate initial conditions for cosmological N-body simulations: minimizing truncation and discreteness errors},
   volume={500},
   ISSN={1365-2966},
   url={http://dx.doi.org/10.1093/mnras/staa3149},
   DOI={10.1093/mnras/staa3149},
   number={1},
   journal={MNRAS},
   publisher={Oxford University Press (OUP)},
   author={Michaux, Michaël and Hahn, Oliver and Rampf, Cornelius and Angulo, Raul E},
   year={2020},
   month=nov, pages={663–683} }

@article{Hahn2021,
   title={Higher order initial conditions for mixed baryon–CDM simulations},
   volume={503},
   ISSN={1365-2966},
   url={http://dx.doi.org/10.1093/mnras/staa3773},
   DOI={10.1093/mnras/staa3773},
   number={1},
   journal={MNRAS},
   publisher={Oxford University Press (OUP)},
   author={Hahn, Oliver and Rampf, Cornelius and Uhlemann, Cora},
   year={2021},
   month=dec, pages={426–445} }

@article{Schaye2023,
   title={The FLAMINGO project: cosmological hydrodynamical simulations for large-scale structure and galaxy cluster surveys},
   volume={526},
   ISSN={1365-2966},
   url={http://dx.doi.org/10.1093/mnras/stad2419},
   DOI={10.1093/mnras/stad2419},
   number={4},
   journal={MNRAS},
   publisher={Oxford University Press (OUP)},
   author={Schaye, Joop and Kugel, Roi and Schaller, Matthieu and Helly, John C and Braspenning, Joey and Elbers, Willem and McCarthy, Ian G and vanDaalen, Marcel P and Vandenbroucke, Bert and Frenk, Carlos S and Kwan, Juliana and Salcido, Jaime and Bahé, Yannick M and Borrow, Josh and Chaikin, Evgenii and Hahn, Oliver and Huško, Filip and Jenkins, Adrian and Lacey, Cedric G and Nobels, Folkert S J},
   year={2023},
   month=aug, pages={4978–5020} }

@article{Han2012,
  title = {Resolving subhaloes’ lives with the Hierarchical Bound-Tracing algorithm: Hierarchical Bound-Tracing algorithm},
  volume = {427},
  ISSN = {1365-2966},
  url = {http://dx.doi.org/10.1111/j.1365-2966.2012.22111.x},
  DOI = {10.1111/j.1365-2966.2012.22111.x},
  number = {3},
  journal = {MNRAS},
  publisher = {Oxford University Press (OUP)},
  author = {Han,  Jiaxin and Jing,  Y. P. and Wang,  Huiyuan and Wang,  Wenting},
  year = {2012},
  month = nov,
  pages = {2437–2449}
}

@article{Jenkins2013,
   title={A new way of setting the phases for cosmological multiscale Gaussian initial conditions},
   volume={434},
   ISSN={1365-2966},
   url={http://dx.doi.org/10.1093/mnras/stt1154},
   DOI={10.1093/mnras/stt1154},
   number={3},
   journal={MNRAS},
   publisher={Oxford University Press (OUP)},
   author={Jenkins, Adrian},
   year={2013},
   month=jul, pages={2094–2120} 
}

@article{Schauer2019,
  title = {The influence of streaming velocities on the formation of the first stars},
  volume = {484},
  ISSN = {1365-2966},
  url = {http://dx.doi.org/10.1093/mnras/stz013},
  DOI = {10.1093/mnras/stz013},
  number = {3},
  journal = {MNRAS},
  publisher = {Oxford University Press (OUP)},
  author = {Schauer,  Anna T P and Glover,  Simon C O and Klessen,  Ralf S and Ceverino,  Daniel},
  year = {2019},
  month = jan,
  pages = {3510–3521}
}

@article{Popa2016,
   title={Gas-rich and gas-poor structures through the stream velocity effect},
   volume={460},
   ISSN={1365-2966},
   url={http://dx.doi.org/10.1093/mnras/stw1045},
   DOI={10.1093/mnras/stw1045},
   number={2},
   journal={MNRAS},
   publisher={Oxford University Press (OUP)},
   author={Popa, Cristina and Naoz, Smadar and Marinacci, Federico and Vogelsberger, Mark},
   year={2016},
   month=may, pages={1625–1639} }

@article{Naoz2012,
   title={SIMULATIONS OF EARLY BARYONIC STRUCTURE FORMATION WITH STREAM VELOCITY. I. HALO ABUNDANCE},
   volume={747},
   ISSN={1538-4357},
   url={http://dx.doi.org/10.1088/0004-637X/747/2/128},
   DOI={10.1088/0004-637x/747/2/128},
   number={2},
   journal={ApJ},
   publisher={American Astronomical Society},
   author={Naoz, Smadar and Yoshida, Naoki and Gnedin, Nickolay Y.},
   year={2012},
   month=feb, pages={128} }

@article{Han2017,
  title = {hbt+: an improved code for finding subhaloes and building merger trees in cosmological simulations},
  volume = {474},
  ISSN = {1365-2966},
  url = {http://dx.doi.org/10.1093/mnras/stx2792},
  DOI = {10.1093/mnras/stx2792},
  number = {1},
  journal = {MNRAS},
  publisher = {Oxford University Press (OUP)},
  author = {Han,  Jiaxin and Cole,  Shaun and Frenk,  Carlos S. and Benitez-Llambay,  Alejandro and Helly,  John},
  year = {2017},
  month = oct,
  pages = {604–617}
}

@article{Rampf2020,
   title={Cosmological perturbations for two cold fluids in ΛCDM},
   volume={503},
   ISSN={1365-2966},
   url={http://dx.doi.org/10.1093/mnras/staa3605},
   DOI={10.1093/mnras/staa3605},
   number={1},
   journal={MNRAS},
   publisher={Oxford University Press (OUP)},
   author={Rampf, Cornelius and Uhlemann, Cora and Hahn, Oliver},
   year={2020},
   month=dec, pages={406–425} }

@article{Bird2020,
   title={More accurate simulations with separate initial conditions for baryons and dark matter},
   volume={2020},
   ISSN={1475-7516},
   url={http://dx.doi.org/10.1088/1475-7516/2020/06/002},
   DOI={10.1088/1475-7516/2020/06/002},
   number={06},
   journal={JCAP},
   publisher={IOP Publishing},
   author={Bird, Simeon and Feng, Yu and Pedersen, Christian and Font-Ribera, Andreu},
   year={2020},
   month=jun, pages={002–002} }

@article{Kugel2023,
  title = {FLAMINGO: calibrating large cosmological hydrodynamical simulations with machine learning},
  volume = {526},
  ISSN = {1365-2966},
  url = {http://dx.doi.org/10.1093/mnras/stad2540},
  DOI = {10.1093/mnras/stad2540},
  number = {4},
  journal = {MNRAS},
  publisher = {Oxford University Press (OUP)},
  author = {Kugel,  Roi and Schaye,  Joop and Schaller,  Matthieu and Helly,  John C and Braspenning,  Joey and Elbers,  Willem and Frenk,  Carlos S and McCarthy,  Ian G and Kwan,  Juliana and Salcido,  Jaime and van Daalen,  Marcel P and Vandenbroucke,  Bert and Bahé,  Yannick M and Borrow,  Josh and Chaikin,  Evgenii and Huško,  Filip and Jenkins,  Adrian and Lacey,  Cedric G and Nobels,  Folkert S J and Vernon,  Ian},
  year = {2023},
  month = oct,
  pages = {6103–6127}
}

@article{Driver2022,
   title={Galaxy And Mass Assembly (GAMA): Data Release 4 and the z < 0.1 total and z < 0.08 morphological galaxy stellar mass functions},
   volume={513},
   ISSN={1365-2966},
   url={http://dx.doi.org/10.1093/mnras/stac472},
   DOI={10.1093/mnras/stac472},
   number={1},
   journal={MNRAS},
   publisher={Oxford University Press (OUP)},
   author={Driver, Simon P and Bellstedt, Sabine and Robotham, Aaron S G and Baldry, Ivan K and Davies, Luke J and Liske, Jochen and Obreschkow, Danail and Taylor, Edward N and Wright, Angus H and Alpaslan, Mehmet and Bamford, Steven P and Bauer, Amanda E and Bland-Hawthorn, Joss and Bilicki, Maciej and Bravo, Matías and Brough, Sarah and Casura, Sarah and Cluver, Michelle E and Colless, Matthew and Conselice, Christopher J and Croom, Scott M and deJong, Jelte and D’Eugenio, Franceso and DePropris, Roberto and Dogruel, Burak and Drinkwater, Michael J and Dvornik, Andrej and Farrow, Daniel J and Frenk, Carlos S and Giblin, Benjamin and Graham, Alister W and Grootes, Meiert W and Gunawardhana, Madusha L P and Hashemizadeh, Abdolhosein and Häußler, Boris and Heymans, Catherine and Hildebrandt, Hendrik and Holwerda, Benne W and Hopkins, Andrew M and Jarrett, Tom H and Heath Jones, D and Kelvin, Lee S and Koushan, Soheil and Kuijken, Konrad and Lara-López, Maritza A and Lange, Rebecca and López-Sánchez, Ángel R and Loveday, Jon and Mahajan, Smriti and Meyer, Martin and Moffett, Amanda J and Napolitano, Nicola R and Norberg, Peder and Owers, Matt S and Radovich, Mario and Raouf, Mojtaba and Peacock, John A and Phillipps, Steven and Pimbblet, Kevin A and Popescu, Cristina and Said, Khaled and Sansom, Anne E and Seibert, Mark and Sutherland, Will J and Thorne, Jessica E and Tuffs, Richard J and Turner, Ryan and vanderwel, Arjen and vanKampen, Eelco and Wilkins, Steve M},
   year={2022},
   month=mar, pages={439–467} }

@article{Borrow2022,
   title={<scp>Sphenix</scp>: smoothed particle hydrodynamics for the next generation of galaxy formation simulations},
   volume={511},
   ISSN={1365-2966},
   url={http://dx.doi.org/10.1093/mnras/stab3166},
   DOI={10.1093/mnras/stab3166},
   number={2},
   journal={MNRAS},
   publisher={Oxford University Press (OUP)},
   author={Borrow, Josh and Schaller, Matthieu and Bower, Richard G and Schaye, Joop},
   year={2022},
   month=nov, pages={2367–2389} }

@article{Ploeckinger2020,
   title={Radiative cooling rates, ion fractions, molecule abundances, and line emissivities including self-shielding and both local and metagalactic radiation fields},
   volume={497},
   ISSN={1365-2966},
   url={http://dx.doi.org/10.1093/mnras/staa2172},
   DOI={10.1093/mnras/staa2172},
   number={4},
   journal={MNRAS},
   publisher={Oxford University Press (OUP)},
   author={Ploeckinger, Sylvia and Schaye, Joop},
   year={2020},
   month=jul, pages={4857–4883} }

@article{Schaye2008,
   title={On the relation between the Schmidt and Kennicutt-Schmidt star formation laws and its implications for numerical simulations: Schmidt and Kennicutt-Schmidt laws},
   volume={383},
   ISSN={1365-2966},
   url={http://dx.doi.org/10.1111/j.1365-2966.2007.12639.x},
   DOI={10.1111/j.1365-2966.2007.12639.x},
   number={3},
   journal={MNRAS},
   publisher={Oxford University Press (OUP)},
   author={Schaye, Joop and Dalla Vecchia, Claudio},
   year={2008},
   month=dec, pages={1210–1222} }

@article{Schaller2024,
   title={Swift: a modern highly parallel gravity and smoothed particle hydrodynamics solver for astrophysical and cosmological applications},
   volume={530},
   ISSN={1365-2966},
   url={http://dx.doi.org/10.1093/mnras/stae922},
   DOI={10.1093/mnras/stae922},
   number={2},
   journal={MNRAS},
   publisher={Oxford University Press (OUP)},
   author={Schaller, Matthieu and Borrow, Josh and Draper, Peter W and Ivkovic, Mladen and McAlpine, Stuart and Vandenbroucke, Bert and Bahé, Yannick and Chaikin, Evgenii and Chalk, Aidan B G and Chan, Tsang Keung and Correa, Camila and van Daalen, Marcel and Elbers, Willem and Gonnet, Pedro and Hausammann, Loïc and Helly, John and Huško, Filip and Kegerreis, Jacob A and Nobels, Folkert S J and Ploeckinger, Sylvia and Revaz, Yves and Roper, William J and Ruiz-Bonilla, Sergio and Sandnes, Thomas D and Uyttenhove, Yolan and Willis, James S and Xiang, Zhen},
   year={2024},
   month=mar, pages={2378–2419} }

@article{Elbers2022,
   title={Higher order initial conditions with massive neutrinos},
   volume={516},
   ISSN={1365-2966},
   url={http://dx.doi.org/10.1093/mnras/stac2365},
   DOI={10.1093/mnras/stac2365},
   number={3},
   journal={MNRAS},
   publisher={Oxford University Press (OUP)},
   author={Elbers, Willem and Frenk, Carlos S and Jenkins, Adrian and Li, Baojiu and Pascoli, Silvia},
   year={2022},
   month=aug, pages={3821–3836} }

@article{Lesgourgues2011,
      title={The Cosmic Linear Anisotropy Solving System (CLASS) I: Overview}, 
      author={Julien Lesgourgues},
      year={2011},
      eprint={1104.2932},
      journal={arXiv preprints},
      archivePrefix={arXiv},
      primaryClass={astro-ph.IM},
      url={https://arxiv.org/abs/1104.2932}, 
}

@article{Katz2025,
  url = {https://arxiv.org/abs/2510.05201},
  author = {Katz,  Harley and Rey,  Martin P. and Cadiou,  Corentin and Agertz,  Oscar and Blaizot,  Jeremy and Cameron,  Alex J. and Choustikov,  Nicholas and Devriendt,  Julien and Hauk,  Uliana and Jones,  Gareth C. and Kimm,  Taysun and Laseter,  Isaac and Martin-Alvarez,  Sergio and Matsumoto,  Kosei and Pearce,  Autumn and Montero,  Francisco Rodríguez and Rosdahl,  Joki and Sanati,  Mahsa and Saxena,  Aayush and Slyz,  Adrianne and Stiskalek,  Richard and Storck,  Anatole and Veenema,  Oscar and Yee,  Wonjae},
  title = {MEGATRON: Reproducing the Diversity of High-Redshift Galaxy Spectra with Cosmological Radiation Hydrodynamics Simulations},
  publisher = {arXiv},
  journal={arXiv preprints},
  pages={arXiv:2510.05201},
  year = {2025},
  copyright = {Creative Commons Attribution 4.0 International}
}

@article{Peroux2020,
   title={The Cosmic Baryon and Metal Cycles},
   volume={58},
   ISSN={1545-4282},
   url={http://dx.doi.org/10.1146/annurev-astro-021820-120014},
   DOI={10.1146/annurev-astro-021820-120014},
   number={1},
   journal={ARA&A},
   publisher={Annual Reviews},
   author={Péroux, Céline and Howk, J. Christopher},
   year={2020},
   month=aug, pages={363–406} 
}

@article{Tuominen2021,
   title={An EAGLE view of the missing baryons},
   volume={646},
   ISSN={1432-0746},
   url={http://dx.doi.org/10.1051/0004-6361/202039221},
   DOI={10.1051/0004-6361/202039221},
   journal={A&A},
   publisher={EDP Sciences},
   author={Tuominen, T. and Nevalainen, J. and Tempel, E. and Kuutma, T. and Wijers, N. and Schaye, J. and Heinämäki, P. and Bonamente, M. and Ganeshaiah Veena, P.},
   year={2021},
   month=feb, pages={A156} }

@article{Yang2022,
   title={Finding the Missing Baryons in the Intergalactic Medium with Localized Fast Radio Bursts},
   volume={940},
   ISSN={2041-8213},
   url={http://dx.doi.org/10.3847/2041-8213/aca145},
   DOI={10.3847/2041-8213/aca145},
   number={2},
   journal={ApJ Letters},
   publisher={American Astronomical Society},
   author={Yang, K. B. and Wu, Q. and Wang, F. Y.},
   year={2022},
   month=nov, pages={L29} }

@article{Gordon2025,
  title = {Conditions for super-Eddington accretion onto the first black holes},
  volume = {537},
  ISSN = {1365-2966},
  url = {http://dx.doi.org/10.1093/mnras/staf054},
  DOI = {10.1093/mnras/staf054},
  number = {2},
  journal = {MNRAS},
  publisher = {Oxford University Press (OUP)},
  author = {Gordon,  Simone T and Smith,  Britton D and Khochfar,  Sadegh and Beckmann,  Ricarda S},
  year = {2025},
  month = jan,
  pages = {674–690}
}

@article{Ni2022,
  title = {The ASTRID simulation: the evolution of supermassive black holes},
  volume = {513},
  ISSN = {1365-2966},
  url = {http://dx.doi.org/10.1093/mnras/stac351},
  DOI = {10.1093/mnras/stac351},
  number = {1},
  journal = {MNRAS},
  publisher = {Oxford University Press (OUP)},
  author = {Ni,  Yueying and Di Matteo,  Tiziana and Bird,  Simeon and Croft,  Rupert and Feng,  Yu and Chen,  Nianyi and Tremmel,  Michael and DeGraf,  Colin and Li,  Yin},
  year = {2022},
  month = feb,
  pages = {670–692}
}

@article{Goulding2023,
  title = {UNCOVER: The Growth of the First Massive Black Holes from JWST/NIRSpec—Spectroscopic Redshift Confirmation of an X-Ray Luminous AGN at z = 10.1},
  volume = {955},
  ISSN = {2041-8213},
  url = {http://dx.doi.org/10.3847/2041-8213/acf7c5},
  DOI = {10.3847/2041-8213/acf7c5},
  number = {1},
  journal = {ApJ Letters},
  publisher = {American Astronomical Society},
  author = {Goulding,  Andy D. and Greene,  Jenny E. and Setton,  David J. and Labbe,  Ivo and Bezanson,  Rachel and Miller,  Tim B. and Atek,  Hakim and Bogdán,  Ákos and Brammer,  Gabriel and Chemerynska,  Iryna and Cutler,  Sam E. and Dayal,  Pratika and Fudamoto,  Yoshinobu and Fujimoto,  Seiji and Furtak,  Lukas J. and Kokorev,  Vasily and Khullar,  Gourav and Leja,  Joel and Marchesini,  Danilo and Natarajan,  Priyamvada and Nelson,  Erica and Oesch,  Pascal A. and Pan,  Richard and Papovich,  Casey and Price,  Sedona H. and van Dokkum,  Pieter and Wang 王,  Bingjie 冰洁 and Weaver,  John R. and Whitaker,  Katherine E. and Zitrin,  Adi},
  year = {2023},
  month = sep,
  pages = {L24}
}

@article{Juodbalis2024,
  title = {A dormant overmassive black hole in the early Universe},
  volume = {636},
  ISSN = {1476-4687},
  url = {http://dx.doi.org/10.1038/s41586-024-08210-5},
  DOI = {10.1038/s41586-024-08210-5},
  number = {8043},
  journal = {Nature},
  publisher = {Springer Science and Business Media LLC},
  author = {Juodžbalis,  Ignas and Maiolino,  Roberto and Baker,  William M. and Tacchella,  Sandro and Scholtz,  Jan and D’Eugenio,  Francesco and Witstok,  Joris and Schneider,  Raffaella and Trinca,  Alessandro and Valiante,  Rosa and DeCoursey,  Christa and Curti,  Mirko and Carniani,  Stefano and Chevallard,  Jacopo and de Graaff,  Anna and Arribas,  Santiago and Bennett,  Jake S. and Bourne,  Martin A. and Bunker,  Andrew J. and Charlot,  Stéphane and Jiang,  Brian and Koudmani,  Sophie and Perna,  Michele and Robertson,  Brant and Sijacki,  Debora and \"{U}bler,  Hannah and Williams,  Christina C. and Willott,  Chris},
  year = {2024},
  month = dec,
  pages = {594–597}
}

@article{Naoz2011,
  title = {The non-linear evolution of baryonic overdensities in the early universe: initial conditions of numerical simulations: Gas rich haloes - sims vs. linear theory},
  ISSN = {0035-8711},
  url = {http://dx.doi.org/10.1111/j.1365-2966.2011.19025.x},
  DOI = {10.1111/j.1365-2966.2011.19025.x},
  journal = {MNRAS},
  publisher = {Oxford University Press (OUP)},
  author = {Naoz,  Smadar and Yoshida,  Naoki and Barkana,  Rennan},
  year = {2011},
  month = jul,
  pages = {no--no}
}

@article{Maiolino2024,
  title = {JADES: The diverse population of infant black holes at 4 &lt; z &lt; 11: Merging,  tiny,  poor,  but mighty},
  volume = {691},
  ISSN = {1432-0746}, 
  url = {http://dx.doi.org/10.1051/0004-6361/202347640},
  DOI = {10.1051/0004-6361/202347640},
  journal = {A&A},
  publisher = {EDP Sciences},
  author = {Maiolino,  Roberto and Scholtz,  Jan and Curtis-Lake,  Emma and Carniani,  Stefano and Baker,  William and de Graaff,  Anna and Tacchella,  Sandro and \"{U}bler,  Hannah and D’Eugenio,  Francesco and Witstok,  Joris and Curti,  Mirko and Arribas,  Santiago and Bunker,  Andrew J. and Charlot,  Stéphane and Chevallard,  Jacopo and Eisenstein,  Daniel J. and Egami,  Eiichi and Ji,  Zhiyuan and Jones,  Gareth C. and Lyu,  Jianwei and Rawle,  Tim and Robertson,  Brant and Rujopakarn,  Wiphu and Perna,  Michele and Sun,  Fengwu and Venturi,  Giacomo and Williams,  Christina C. and Willott,  Chris},
  year = {2024},
  month = nov,
  pages = {A145}
}

@article{Kocevski2025,
  title = {The Rise of Faint,  Red Active Galactic Nuclei at z &gt; 4: A Sample of Little Red Dots in the JWST Extragalactic Legacy Fields},
  volume = {986},
  ISSN = {1538-4357},
  url = {http://dx.doi.org/10.3847/1538-4357/adbc7d},
  DOI = {10.3847/1538-4357/adbc7d},
  number = {2},
  journal = {ApJ},
  publisher = {American Astronomical Society},
  author = {Kocevski,  Dale D. and Finkelstein,  Steven L. and Barro,  Guillermo and Taylor,  Anthony J. and Calabrò,  Antonello and Laloux,  Brivael and Buchner,  Johannes and Trump,  Jonathan R. and Leung,  Gene C. K. and Yang,  Guang and Dickinson,  Mark and Pérez-González,  Pablo G. and Pacucci,  Fabio and Inayoshi,  Kohei and Somerville,  Rachel S. and McGrath,  Elizabeth J. and Akins,  Hollis B. and Bagley,  Micaela B. and Bowler,  Rebecca A.A. and Bisigello,  Laura and Carnall,  Adam and Casey,  Caitlin M. and Cheng,  Yingjie and Cleri,  Nikko J. and Costantin,  Luca and Cullen,  Fergus and Davis,  Kelcey and Donnan,  Callum T. and Dunlop,  James S. and Ellis,  Richard S. and Ferguson,  Henry C. and Fujimoto,  Seiji and Fontana,  Adriano and Giavalisco,  Mauro and Grazian,  Andrea and Grogin,  Norman A. and Hathi,  Nimish P. and Hirschmann,  Michaela and Huertas-Company,  Marc and Holwerda,  Benne W. and Illingworth,  Garth and Juneau,  Stéphanie and Kartaltepe,  Jeyhan S. and Koekemoer,  Anton M. and Li,  Wenxiu and Lucas,  Ray A. and Magee,  Dan and Mason,  Charlotte and McLeod,  Derek J. and McLure,  Ross J. and Napolitano,  Lorenzo and Papovich,  Casey and Pirzkal,  Nor and Rodighiero,  Giulia and Santini,  Paola and Wilkins,  Stephen M. and Yung,  L. Y. Aaron},
  year = {2025},
  month = jun,
  pages = {126}
}

@article{Li2025,
  title = {Tip of the Iceberg: Overmassive Black Holes at 4 &lt; z &lt; 7 Found by JWST Are Not Inconsistent with the Local 
	    <mml:math xmlns:mml="http://www.w3.org/1998/Math/MathML">
                     <mml:msub>
                        <mml:mi>M</mml:mi>
                        <mml:mi>BH</mml:mi>
                     </mml:msub>
                     <mml:mo>-</mml:mo>
                     <mml:msub>
                        <mml:mi>M</mml:mi>
                        <mml:mo>⋆</mml:mo>
                     </mml:msub>
                  </mml:math>
	   Relation},
  volume = {981},
  ISSN = {1538-4357},
  url = {http://dx.doi.org/10.3847/1538-4357/ada603},
  DOI = {10.3847/1538-4357/ada603},
  number = {1},
  journal = {ApJ},
  publisher = {American Astronomical Society},
  author = {Li,  Junyao and Silverman,  John D. and Shen,  Yue and Volonteri,  Marta and Jahnke,  Knud and Zhuang,  Ming-Yang and Scoggins,  Matthew T. and Ding,  Xuheng and Harikane,  Yuichi and Onoue,  Masafusa and Tanaka,  Takumi S.},
  year = {2025},
  month = feb,
  pages = {19}
}

@article{Pacucci2023,
  title = {JWST CEERS and JADES Active Galaxies at z = 4–7 Violate the Local M–M⋆ Relation at 3σ: Implications for Low-mass Black Holes and Seeding Models},
  volume = {957},
  ISSN = {2041-8213},
  url = {http://dx.doi.org/10.3847/2041-8213/ad0158},
  DOI = {10.3847/2041-8213/ad0158},
  number = {1},
  journal = {ApJ Letters},
  publisher = {American Astronomical Society},
  author = {Pacucci,  Fabio and Nguyen,  Bao and Carniani,  Stefano and Maiolino,  Roberto and Fan,  Xiaohui},
  year = {2023},
  month = oct,
  pages = {L3}
}

@article{Regan2024,
   title={Massive Black Hole Seeds},
   volume={7},
   ISSN={2565-6120},
   url={http://dx.doi.org/10.33232/001c.123239},
   DOI={10.33232/001c.123239},
   journal={The Open Journal of Astrophysics},
   publisher={Maynooth University},
   author={Regan, John and Volonteri, Marta},
   year={2024},
   month=sep }

@article{Inayoshi2020,
   title={The Assembly of the First Massive Black Holes},
   volume={58},
   ISSN={1545-4282},
   url={http://dx.doi.org/10.1146/annurev-astro-120419-014455},
   DOI={10.1146/annurev-astro-120419-014455},
   number={1},
   journal={ARA&A},
   publisher={Annual Reviews},
   author={Inayoshi, Kohei and Visbal, Eli and Haiman, Zoltán},
   year={2020},
   month=aug, pages={27–97} }

@article{Greene2024,
  title = {UNCOVER Spectroscopy Confirms the Surprising Ubiquity of Active Galactic Nuclei in Red Sources at z &gt; 5},
  volume = {964},
  ISSN = {1538-4357},
  url = {http://dx.doi.org/10.3847/1538-4357/ad1e5f},
  DOI = {10.3847/1538-4357/ad1e5f},
  number = {1},
  journal = {ApJ},
  publisher = {American Astronomical Society},
  author = {Greene,  Jenny E. and Labbe,  Ivo and Goulding,  Andy D. and Furtak,  Lukas J. and Chemerynska,  Iryna and Kokorev,  Vasily and Dayal,  Pratika and Volonteri,  Marta and Williams,  Christina C. and Wang 王,  Bingjie 冰洁 and Setton,  David J. and Burgasser,  Adam J. and Bezanson,  Rachel and Atek,  Hakim and Brammer,  Gabriel and Cutler,  Sam E. and Feldmann,  Robert and Fujimoto,  Seiji and Glazebrook,  Karl and de Graaff,  Anna and Khullar,  Gourav and Leja,  Joel and Marchesini,  Danilo and Maseda,  Michael V. and Matthee,  Jorryt and Miller,  Tim B. and Naidu,  Rohan P. and Nanayakkara,  Themiya and Oesch,  Pascal A. and Pan,  Richard and Papovich,  Casey and Price,  Sedona H. and van Dokkum,  Pieter and Weaver,  John R. and Whitaker,  Katherine E. and Zitrin,  Adi},
  year = {2024},
  month = mar,
  pages = {39}
}

@article{Matthee2024,
  title = {Little Red Dots: An Abundant Population of Faint Active Galactic Nuclei at z ∼ 5 Revealed by the EIGER and FRESCO JWST Surveys},
  volume = {963},
  ISSN = {1538-4357},
  url = {http://dx.doi.org/10.3847/1538-4357/ad2345},
  DOI = {10.3847/1538-4357/ad2345},
  number = {2},
  journal = {ApJ},
  publisher = {American Astronomical Society},
  author = {Matthee,  Jorryt and Naidu,  Rohan P. and Brammer,  Gabriel and Chisholm,  John and Eilers,  Anna-Christina and Goulding,  Andy and Greene,  Jenny and Kashino,  Daichi and Labbe,  Ivo and Lilly,  Simon J. and Mackenzie,  Ruari and Oesch,  Pascal A. and Weibel,  Andrea and Wuyts,  Stijn and Xiao,  Mengyuan and Bordoloi,  Rongmon and Bouwens,  Rychard and van Dokkum,  Pieter and Illingworth,  Garth and Kramarenko,  Ivan and Maseda,  Michael V. and Mason,  Charlotte and Meyer,  Romain A. and Nelson,  Erica J. and Reddy,  Naveen A. and Shivaei,  Irene and Simcoe,  Robert A. and Yue,  Minghao},
  year = {2024},
  month = mar,
  pages = {129}
}

@article{Labbe2023,
  title = {A population of red candidate massive galaxies ~600 Myr after the Big Bang},
  volume = {616},
  ISSN = {1476-4687},
  url = {http://dx.doi.org/10.1038/s41586-023-05786-2},
  DOI = {10.1038/s41586-023-05786-2},
  number = {7956},
  journal = {Nature},
  publisher = {Springer Science and Business Media LLC},
  author = {Labbé,  Ivo and van Dokkum,  Pieter and Nelson,  Erica and Bezanson,  Rachel and Suess,  Katherine A. and Leja,  Joel and Brammer,  Gabriel and Whitaker,  Katherine and Mathews,  Elijah and Stefanon,  Mauro and Wang,  Bingjie},
  year = {2023},
  month = feb,
  pages = {266–269}
}

@article{Naoz2005,
  title = {Growth of linear perturbations before the era of the first galaxies},
  volume = {362},
  ISSN = {1365-2966},
  url = {http://dx.doi.org/10.1111/j.1365-2966.2005.09385.x},
  DOI = {10.1111/j.1365-2966.2005.09385.x},
  number = {3},
  journal = {MNRAS},
  publisher = {Oxford University Press (OUP)},
  author = {Naoz,  S. and Barkana,  R.},
  year = {2005},
  month = sep,
  pages = {1047–1053}
}

@article{Barkana2011,
  title = {Scale-dependent bias of galaxies from baryonic acoustic oscillations: Scale-dependent bias from BAOs},
  volume = {415},
  ISSN = {0035-8711},
  url = {http://dx.doi.org/10.1111/j.1365-2966.2011.18922.x},
  DOI = {10.1111/j.1365-2966.2011.18922.x},
  number = {4},
  journal = {MNRAS},
  publisher = {Oxford University Press (OUP)},
  author = {Barkana,  Rennan and Loeb,  Abraham},
  year = {2011},
  month = jun,
  pages = {3113–3118}
}

@article{Pacucci2024,
   title={The Redshift Evolution of the M•–M⋆ Relation for JWST’s Supermassive Black Holes at z &gt; 4},
   volume={964},
   ISSN={1538-4357},
   url={http://dx.doi.org/10.3847/1538-4357/ad3044},
   DOI={10.3847/1538-4357/ad3044},
   number={2},
   journal={ApJ},
   publisher={American Astronomical Society},
   author={Pacucci, Fabio and Loeb, Abraham},
   year={2024},
   month=mar, pages={154} }

@article{Juodzbalis2023,
   title={EPOCHS VII: discovery of high-redshift (6.5 < z < 12) AGN candidates in JWST ERO and PEARLS data},
   volume={525},
   ISSN={1365-2966},
   url={http://dx.doi.org/10.1093/mnras/stad2396},
   DOI={10.1093/mnras/stad2396},
   number={1},
   journal={MNRAS},
   publisher={Oxford University Press (OUP)},
   author={Juodžbalis, Ignas and Conselice, Christopher J and Singh, Maitrayee and Adams, Nathan and Ormerod, Katherine and Harvey, Thomas and Austin, Duncan and Volonteri, Marta and Cohen, Seth H and Jansen, Rolf A and Summers, Jake and Windhorst, Rogier A and D’Silva, Jordan C J and Koekemoer, Anton M and Coe, Dan and Driver, Simon P and Frye, Brenda and Grogin, Norman A and Marshall, Madeline A and Nonino, Mario and Pirzkal, Nor and Robotham, Aaron and Jr, Russell E Ryan and Ortiz III, Rafael and Tompkins, Scott and Willmer, Christopher N A and Yan, Haojing},
   year={2023},
   month=aug, pages={1353–1364} }

@article{Wang2023b,
   title={An efficient and robust method to estimate halo concentration based on the method of moments},
   volume={527},
   ISSN={1365-2966},
   url={http://dx.doi.org/10.1093/mnras/stad3927},
   DOI={10.1093/mnras/stad3927},
   number={4},
   journal={MNRAS},
   publisher={Oxford University Press (OUP)},
   author={Wang, Kai and Mo, H J and Chen, Yangyao and Schaye, Joop},
   year={2023},
   month=dec, pages={10760–10776} }

@article{Dave2019,
   title={simba: Cosmological simulations with black hole growth and feedback},
   volume={486},
   ISSN={1365-2966},
   url={http://dx.doi.org/10.1093/mnras/stz937},
   DOI={10.1093/mnras/stz937},
   number={2},
   journal={MNRAS},
   publisher={Oxford University Press (OUP)},
   author={Davé, Romeel and Anglés-Alcázar, Daniel and Narayanan, Desika and Li, Qi and Rafieferantsoa, Mika H and Appleby, Sarah},
   year={2019},
   month=apr, pages={2827–2849} }

@article{Kaviraj2017,
   title={The Horizon-AGN simulation: evolution of galaxy properties over cosmic time},
   ISSN={1365-2966},
   url={http://dx.doi.org/10.1093/mnras/stx126},
   DOI={10.1093/mnras/stx126},
   journal={MNRAS},
   publisher={Oxford University Press (OUP)},
   author={Kaviraj, S. and Laigle, C. and Kimm, T. and Devriendt, J. E. G. and Dubois, Y. and Pichon, C. and Slyz, A. and Chisari, E. and Peirani, S.},
   year={2017},
   month=jan, pages={stx126} }

@article{Rosdahl2018,
   title={The SPHINX Cosmological Simulations of the First Billion Years: the Impact of Binary Stars on Reionization★},
   ISSN={1365-2966},
   url={http://dx.doi.org/10.1093/mnras/sty1655},
   DOI={10.1093/mnras/sty1655},
   journal={MNRAS},
   publisher={Oxford University Press (OUP)},
   author={Rosdahl, Joakim and Katz, Harley and Blaizot, Jérémy and Kimm, Taysun and Michel-Dansac, Léo and Garel, Thibault and Haehnelt, Martin and Ocvirk, Pierre and Teyssier, Romain},
   year={2018},
   month=jun }

@article{Latif2016,
  title = {Formation of Supermassive Black Hole Seeds},
  volume = {33},
  ISSN = {1448-6083},
  url = {http://dx.doi.org/10.1017/pasa.2016.41},
  DOI = {10.1017/pasa.2016.41},
  journal = {PASA},
  publisher = {Cambridge University Press (CUP)},
  author = {Latif,  Muhammad A. and Ferrara,  Andrea},
  year = {2016}
}

@article{Barkana2005,
  title = {Probing the epoch of early baryonic infall through 21-cm fluctuations},
  volume = {363},
  ISSN = {1745-3925},
  url = {http://dx.doi.org/10.1111/j.1745-3933.2005.00079.x},
  DOI = {10.1111/j.1745-3933.2005.00079.x},
  number = {1},
  journal = {MNRAS Letters},
  publisher = {Oxford University Press (OUP)},
  author = {Barkana,  R. and Loeb,  A.},
  year = {2005},
  month = oct,
  pages = {L36–L40}
}

@article{Naoz2007,
   title={The formation and gas content of high-redshift galaxies and minihaloes},
   volume={377},
   ISSN={1365-2966},
   url={http://dx.doi.org/10.1111/j.1365-2966.2007.11636.x},
   DOI={10.1111/j.1365-2966.2007.11636.x},
   number={2},
   journal={MNRAS},
   publisher={Oxford University Press (OUP)},
   author={Naoz, S. and Barkana, R.},
   year={2007},
   month=may, pages={667–676} }

@article{Lee2024,
  title = {Effective equation of state of a radiatively cooling gas: Self-similar solution of spherical collapse},
  volume = {684},
  ISSN = {1432-0746},
  url = {http://dx.doi.org/10.1051/0004-6361/202346533},
  DOI = {10.1051/0004-6361/202346533},
  journal = {A&A},
  publisher = {EDP Sciences},
  author = {Lee,  Yueh-Ning},
  year = {2024},
  month = apr,
  pages = {A48}
}

@article{Devecchi2012,
   title={High-redshift formation and evolution of central massive objects - II. The census of BH seeds: Formation of CMOs-II},
   volume={421},
   ISSN={0035-8711},
   url={http://dx.doi.org/10.1111/j.1365-2966.2012.20406.x},
   DOI={10.1111/j.1365-2966.2012.20406.x},
   number={2},
   journal={MNRAS},
   publisher={Oxford University Press (OUP)},
   author={Devecchi, B. and Volonteri, M. and Rossi, E. M. and Colpi, M. and Portegies Zwart, S.},
   year={2012},
   month=feb, pages={1465–1475} }

@article{Yue2014,
   title={The brief era of direct collapse black hole formation},
   volume={440},
   ISSN={1365-2966},
   url={http://dx.doi.org/10.1093/mnras/stu351},
   DOI={10.1093/mnras/stu351},
   number={2},
   journal={MNRAS},
   publisher={Oxford University Press (OUP)},
   author={Yue, Bin and Ferrara, Andrea and Salvaterra, Ruben and Xu, Yidong and Chen, Xuelei},
   year={2014},
   month=mar, pages={1263–1273} }

@article{LesgourguesTram2011,
   title={The Cosmic Linear Anisotropy Solving System (CLASS) IV: efficient implementation of non-cold relics},
   volume={2011},
   ISSN={1475-7516},
   url={http://dx.doi.org/10.1088/1475-7516/2011/09/032},
   DOI={10.1088/1475-7516/2011/09/032},
   number={09},
   journal={JCAP},
   publisher={IOP Publishing},
   author={Lesgourgues, Julien and Tram, Thomas},
   year={2011},
   month=sep, pages={032–032} }

@article{Chaikin2022,
   title={The importance of the way in which supernova energy is distributed around young stellar populations in simulations of galaxies},
   volume={514},
   ISSN={1365-2966},
   url={http://dx.doi.org/10.1093/mnras/stac1132},
   DOI={10.1093/mnras/stac1132},
   number={1},
   journal={MNRAS},
   publisher={Oxford University Press (OUP)},
   author={Chaikin, Evgenii and Schaye, Joop and Schaller, Matthieu and Bahé, Yannick M and Nobels, Folkert S J and Ploeckinger, Sylvia},
   year={2022},
   month=apr, pages={249–264} }

@article{Chaikin2023,
   title={A thermal–kinetic subgrid model for supernova feedback in simulations of galaxy formation},
   volume={523},
   ISSN={1365-2966},
   url={http://dx.doi.org/10.1093/mnras/stad1626},
   DOI={10.1093/mnras/stad1626},
   number={3},
   journal={MNRAS},
   publisher={Oxford University Press (OUP)},
   author={Chaikin, Evgenii and Schaye, Joop and Schaller, Matthieu and Benítez-Llambay, Alejandro and Nobels, Folkert S J and Ploeckinger, Sylvia},
   year={2023},
   month=jun, pages={3709–3731} }

@article{McGibbon2025,
   title={SOAP: A Python Package for Calculating the Properties of Galaxies and Halos Formed in Cosmological Simulations},
   volume={10},
   ISSN={2475-9066},
   url={http://dx.doi.org/10.21105/joss.08252},
   DOI={10.21105/joss.08252},
   number={111},
   journal={JOSS},
   publisher={The Open Journal},
   author={McGibbon, Robert and Helly, John C. and Schaye, Joop and Schaller, Matthieu and Vandenbroucke, Bert},
   year={2025},
   month=jul, pages={8252} }

@article{Booth2009,
   title={Cosmological simulations of the growth of supermassive black holes and feedback from active galactic nuclei: method and tests},
   volume={398},
   ISSN={1365-2966},
   url={http://dx.doi.org/10.1111/j.1365-2966.2009.15043.x},
   DOI={10.1111/j.1365-2966.2009.15043.x},
   number={1},
   journal={MNRAS},
   publisher={Oxford University Press (OUP)},
   author={Booth, C. M. and Schaye, Joop},
   year={2009},
   month=sep, pages={53–74} }

@article{Press1982,
  title = {How to identify and weigh virialized clusters of galaxies in a complete redshift catalog},
  volume = {259},
  ISSN = {1538-4357},
  url = {http://dx.doi.org/10.1086/160183},
  DOI = {10.1086/160183},
  journal = {ApJ},
  publisher = {American Astronomical Society},
  author = {Press,  W. H. and Davis,  M.},
  year = {1982},
  month = aug,
  pages = {449}
}


\appendix
\section{The time evolution of Baryons and CDM in an Einstein de-Sitter Universe after Baryon--Photon Decoupling}
\label{appendix:simple_argument}

In \cref{subsec:decomp}, we showed that the isocurvature overdensity field, $\delta_{\rm{bc}}$, is strongly anti-correlated with the matter overdensity field, $\delta_{\rm m}$. This behaviour can be understood by specifying the conditions of the baryon and CDM fluids at baryon-photon decoupling and tracing their subsequent evolution. While the coupled baryon--CDM system generally requires a numerical solution, our aim is to construct a simple theoretical model to capture the origin of this anti-correlation on the scales relevant to structure formation. To this end, we consider the time evolution of the two fluids in an Einstein de--Sitter (EdS) Universe in which we ignore the effects of radiation, dark energy, and neutrinos.

We begin by introducing the baryon and CDM overdensity fields, $\delta_{\rm b}(\mathbf{x},\,t)$ and $\delta_{\rm c}(\mathbf{x},\,t)$, defined at comoving position $\mathbf{x}$ and time $t$. We neglect the self-gravity and pressure of baryons, assuming they respond to but do not contribute towards the gravitational potential. This approximation is valid shortly after baryon–photon decoupling on the scales of interest: photon drag had suppressed baryonic perturbations ($\delta_{\rm b} \ll \delta_{\rm c}$), so that $\delta_{\rm m}$, and hence the gravitational potential, is sourced predominantly by $\delta_{\rm c}$.

In this EdS Universe, $\delta_{\rm c}(\mathbf{x},\,t)$ is described by the following second-order differential equation (e.g. \citealt{Weinberg2008}):
\begin{equation}
  \ddot\delta_{\rm c}(\mathbf{x},\,t)
  + \frac{4}{3t}\,\dot\delta_{\rm c}(\mathbf{x},\,t)
  - \frac{2}{3t^2}\,\delta_{\rm c}(\mathbf{x},\,t)
  \;=\;0,
\end{equation}
The growing-mode solution for $\delta_{\rm c}(\mathbf{x},\,t)$ and its time derivative $\dot{\delta}_{\rm c}(\mathbf{x},\,t)$ are given by
\begin{equation}
\label{eq:growing}
  \delta_{\rm c}(\mathbf{x},\,t)
  \;=\;
  A(\mathbf{x})\left(\frac{t}{t_{\rm dec}}\right)^{2/3},
  \qquad
  \dot{\delta}_{\rm c}(\mathbf{x},\,t)
  \;=\;
  \frac{2}{3t}\,\delta_{\rm c}(\mathbf{x},\,t),
\end{equation}
where $t_{\rm dec}$ denotes the epoch of baryon–photon decoupling and $A(\mathbf{x})$ is a dimensionless function defined such that $A(\mathbf{x})=\delta_{\rm c}(\mathbf{x},\,t=t_{\rm dec})$. $\delta_{\rm b}(\mathbf{x},\,t)$ obeys a similar but inhomogeneous differential equation,
\begin{equation}
  \ddot\delta_{\rm b}(\mathbf{x},\,t)
  + \frac{4}{3t}\,\dot\delta_{\rm b}(\mathbf{x},\,t)
 -\frac{2}{3t^2}\,\delta_{\rm c}(\mathbf{x},\,t)=0,
\end{equation}
and substituting \cref{eq:growing} gives
\begin{equation}
  \ddot\delta_{\rm b}(\mathbf{x},\,t) + \frac{4}{3t}\,\dot\delta_{\rm b}(\mathbf{x},\,t)
 -\frac{2}{3}A(\mathbf{x})\,t^{-4/3}t_{\rm dec}^{-2/3}=0.
\end{equation}
Its general solution can be written as
\begin{equation}
\delta_{\rm b}(\mathbf{x},\,t) 
= \delta_{\rm c}(\mathbf{x},\,t) +
C(\mathbf{x})\left(\frac{t}{t_{\rm dec}}\right)^{-1/3} + D(\mathbf{x}),
\qquad t\ge t_{\rm dec},
\end{equation}
where $C(\mathbf{x})$ and $D(\mathbf{x})$ are dimensionless functions.

At decoupling $\delta_{\rm b}(\mathbf{x},\,t)$ is effectively smooth and nearly static, allowing us to impose the following initial conditions
\begin{equation}
  \delta_{\rm b}(\mathbf{x},\,t=t_{\rm dec})=0,
  \qquad
  \dot\delta_{\rm b}(\mathbf{x},\,t=t_{\rm dec})=0,
\end{equation}
which fixes the free functions
\begin{equation}
C(\mathbf{x})=2\,\delta_{\rm c}(\mathbf{x},\,t=t_{\rm dec}), 
\qquad
D(\mathbf{x})=-3\,\delta_{\rm c}(\mathbf{x},\,t=t_{\rm dec}), 
\end{equation}
Combining these boundary conditions with the fact that $\delta_{\rm c} = \delta_{\rm m}$ at decoupling, we find that the isocurvature overdensity field,
\begin{equation}
\delta_{\rm{bc}}(\mathbf{x},\,t) \equiv \delta_{\rm b}(\mathbf{x},\,t) - \delta_{\rm c}(\mathbf{x},\,t), 
\end{equation}
is given by
\begin{equation}
\delta_{\rm{bc}}(\mathbf{x},\,t) 
=-3\,\delta_{\rm m}(\mathbf{x},\,t=t_{\rm dec})\left(1-\frac{2}{3}\,\left(\frac{t}{t_{\rm dec}}\right)^{-1/3}\right),
\,t\ge t_{\rm dec}.
\end{equation}
This expression shows that the amplitude of $\delta_{\rm{bc}}$ is fully determined by the amplitude of $\delta_{\rm{m}}$ at decoupling. In the asymptotic time limit,
\begin{equation}
\label{eq:asymptote}
\delta_{\rm{bc}}(\mathbf{x},\,t\to\infty) = -3\,\delta_{\rm m}(\mathbf{x},\,t=t_{\rm dec}).
\end{equation}
Within this model $\delta_{\rm{bc}}$ is anti-parallel to $\delta_{\rm m}$, such that the two fields are perfectly anti-correlated. This derivation captures the essential origin of the anti-correlation and is consistent with the more general analyses of \citet{Rampf2020} and \citet{Hahn2021}, who solve the coupled baryon–CDM system in a $\Lambda$CDM Universe under the assumption of negligible baryon temperature.

We test the accuracy of this asymptotic model by using it to predict the RMS amplitude of $\delta_{\rm{bc}}$ at $z=0$, combining \cref{eq:asymptote} with the RMS amplitude of $\delta_{\rm m}$ at decoupling, computed using the Einstein–Boltzmann code \texttt{CLASS}. The prediction is then compared directly with the RMS amplitude of $\delta_{\rm{bc}}$ at $z=0$ obtained from \texttt{CLASS}. As in \cref{fig:density_fields}, the density fields are smoothed on the Lagrangian scale corresponding to a $10^{11}\,\rm M_\odot$ halo prior to evaluating the RMS. The asymptotic model yields $\sigma_{\textrm{RMS}}(\delta_{\rm{bc}}) = 0.0132$, while \texttt{CLASS} gives $\sigma_{\textrm{RMS}}(\delta_{\rm{bc}}) = 0.0147$, a relative difference of $10\%$. This close agreement demonstrates that the theoretical model captures the essential physics governing the time evolution of the baryon–CDM isocurvature mode and provides a quantitatively reliable estimate of its late time amplitude on the scales relevant for halo formation.

\bsp	
\label{lastpage}
\end{document}